\documentclass[a4paper,fleqn,usenatbib,useAMS,pdfpagelabels=false,colorlinks=true,allcolors=blue]{mnras}
\usepackage[english]{babel}
\usepackage[utf8x]{inputenc}
\usepackage[T1]{fontenc}
\usepackage{natbib}
\usepackage{subfigure}

\usepackage{amsmath}
\usepackage{txfonts}
\usepackage{mathptmx}
\usepackage{graphicx}
\usepackage{epsfig}
\usepackage{amssymb}
\usepackage{color}

\title{Analyzing the gamma-ray activity of neutrino emitter candidates: comparing TXS 0506+056 with other blazars}
\author[Antonio Marinelli et al.]{
Antonio Marinelli,$^{1}$\thanks{E-mail:antonio.marinelli@pi.infn.it}
J.Rodrigo Sacahui,$^{2}$\thanks{E-mail:jrsacahui@profesor.usac.edu.gt}
Ankur Sharma,$^{3}$\thanks{E-mail:ankur.sharma@physics.uu.se}
and Mabel Osorio-Archila$^{4}$\thanks{E-mail:jmosorio@astro.unam.mx}
\\
\\
$^{1}$Instituto Nazionale di Fisica Nucleare, Sezione di Napoli, Complesso Universitario di Monte S. Angelo, Via Cintia ed.G, Napoli, 80126 Italy \\
$^{2}$Instituto de Investigaci\'on en Ciencias F\'isicas y Matem\'aticas, USAC, Ciudad Universitaria, Zona 12, Guatemala\\
$^{3}$Dept. of Physics and Astronomy, Uppsala University, Box 516, SE-75120 Uppsala, Sweden\\
$^{4}$Instituto de Astronomía, UNAM, Circuito de la Investigaci\'on Cient\'ifica, Ciudad Universitaria, 04510 M\'exico City, M\'exico
}
\begin{document}
\maketitle

\begin{abstract}
On 22nd of September 2017 the IceCube collaboration sent an alert for an EHE (Extreme High Energy) event, corresponding to the reconstruction of a muonic neutrino (IC-170922A) with energy $\sim$290 TeV. A multi-wavelength follow-up campaign associated this neutrino event with a gamma-ray flaring state of the BL Lac TXS 0506+056 located at \textit{z}=0.3365. From the same position of the sky a muonic neutrino excess is observed in a time window of 110 days around $13^{th}$ of December 2014. These observations together suggest TXS 0506+056 as a possible neutrino emitter. We report here a long term gamma-ray monitoring of this source and we compare it with other blazars spatially correlated with astrophysical muonic neutrino events observed by IceCube. We characterize the most significant gamma-ray flares of the blazars in the sample and introduce the blazar duty cycle as an important parameter to be considered when assessing a possible neutrino counterpart.
For the selected blazars we show the expected neutrino flux variability with different time binning assuming the gamma-rays observed by Fermi-LAT as a product of a leptohadronic emission scenario. The neutrino expectations from the blazar sample are then compared with the IceCube discovery flux searching for the optimal time bin in a multi-messenger context. This analysis indicates that the detection of a single flare represents a challenge for a single cubic kilometer detector, underlining the importance of increasing the size of current neutrino telescopes and obtaining a good monitoring of the entire sky through a global neutrino network.
\end{abstract}


\section{Introduction}
Since November 2013, when the IceCube collaboration tagged the first two ``extraterrestrial" high energy neutrino events (``Bert" and ``Ernie")~\citep{2013PhRvL.111b1103A}, the observed astrophysical neutrino signal has exceeded hundreds of events collected in different samples~\citep{2016ApJ...833....3A,2017arXiv171001191I,2019ICRC...36.1004S,2019ICRC...36.1017S}. The spatial distribution of the events do not show particular accumulations in the skymap with a significance greater than 3$\sigma$~\citep{2017ApJ...835..151A,2019ApJ...886...23X}. The spectral energy distribution (SED) of the muonic neutrino events observed from the Northern hemisphere resulted in a harder index ($\alpha\sim 2.2$), than the SED of the full sky High Energy Starting Events (HESE) observed in 7 years~\citep{2017arXiv171001191I}. Recent analyses by IceCube and ANTARES experiments~\citep{2017ApJ...849...67A,2017PhRvD..96f2001A} posed upper limits on the diffuse Galactic emission~\citep{2015ApJ...815L..25G} showing that this contribution cannot exceed $8.5\%$ of the full sky measured neutrino flux. A remaining viable solution for a global neutrino SED description is represented by the existence of different extragalactic populations of neutrino emitters~\citep{2017ApJ...836...47B}, with each of them having a major contribution at different energy range. 
However it is not clear which class of extragalactic sources contributes more to the remaining $\sim 90\%$ of the observed astrophysical neutrino flux~\citep{2014PhRvD..90d3005A,2016PhRvD..94j3006M,2018ApJ...863...64T}. \\
An answer to this question can be obtained through the electromagnetic follow-up of the most energetic IceCube neutrino events.
In 2016 the IceCube collaboration started a public alert program~\citep{2017APh....92...30A}, sending out alerts in real-time whenever an Extreme High Energy (EHE)~\citep{PhysRevD.89.062007,Ishihara_2016} or a High Energy Starting Event (HESE~\citep{2014PhRvL.113j1101A,2017arXiv171001191I}) with a good angular resolution is reconstructed. Up till now, 23 of these alerts have been sent out by the IceCube experiment~\citep{eaat1378} to other astrophysical observatories which can cover the 
electromagnetic sky from radio to gamma-ray frequencies~\citep{2013APh....45...56S}.\\ The first positive follow-up was obtained with the alert sent on $22^{nd}$ of September 2017 when Fermi Large Area Telescope (LAT)~\citep{atwood09} observed the BL Lac TXS 0506+056~\citep{10.1093/mnrasl/slz011} in a flaring state inside the small solid angle associated with an IceCube EHE event (IC-170922A) direction, with energy $\sim$ 290 TeV~\citep{eaat1378}. For the same source, the MAGIC gamma-ray telescope also observed high-energy activity between 80-400 GeV between $28^{th}$ September to $4^{th}$ of October~\citep{2018ApJ_MAGIC}. An unblinded analysis of this BL Lac by IceCube using the data collected in the last 8 years showed a 3.5$\sigma$ excess in December 2014 in the sample of total reconstructed low energy muonic neutrino events~\citep{2018Sci...361..147I}. Unfortunately in this case, Fermi-LAT data did not show a gamma-ray counterpart activity. Different leptohadronic models of TXS 0506+056 were used to describe the state of this source around the muonic neutrino excess of 2014-15~\citep{2018ApJ...865..124M,2019ApJ...881...46R,2019ApJ...874L...9H,2019ApJ...874L..29R} and around the $22^{nd}$ September alert~\citep{2018ApJ...866..109S,2019MNRAS_Zech,2019PhRvD..99f3008L}, to account for the observed gamma-ray and neutrino events. While a unique multi-messenger scenario explaining both neutrino observations seems unviable, a link with the electromagnetic emission seems possible at different energy ranges. To associate the 2014-15 low energy neutrino flare with the absence of enhanced gamma-ray activity, we should invoke the absorption of GeV photons in the source environment and eventually Compton-supported cascades~\citep{2019ApJ...881...46R}. To such a degree, the connection between the 2017 EHE neutrino event, the flaring GeV activity observed by Fermi-LAT and the X-ray observations of NUSTAR~\citep{2018ApJ-Murase,2019ApJ...886...23X} seem non trivial.\\ Here we use the leptohadronic scenario of~\citep{2018ApJ-Murase} to link the Fermi-LAT measurements of TXS 0506+056 (during the 2017 flare) and the expected very-high-energy (VHE) neutrino flux. Additionally, following the model of~\citep{2015MNRAS.448.2412P}  we study possible hadronic emission from a sample of blazars spatially connected with astrophysical IceCube events using a one-zone leptohadronic model to connect the VHE neutrino events with the Fermi-LAT data. In particular we consider the flaring GeV activity observed by Fermi-LAT as the synchrotron emission of charged-pion cascades~\citep{2018MNRAS.480..192P}.
For the sample selection we look for the neutrino track-like events with a reconstruction error $< 1.5^{\circ}$, being part of EHE, HESE and northern hemisphere muonic neutrino~\citep{2016ApJ...833....3A} catalogs and we search for the VHE emitters in the Fermi-LAT 3FGL~\citep{Acero15} and 3FHL~\citep{2017ApJS..232...18A} catalogs spatially connected with the reconstructed astrophysical neutrino events. With these criteria, we construct a sample of blazars that are candidates for VHE neutrino emission. The daily gamma-ray variability and luminosity of these sources~\citep{Urry95,Wills92,1998MNRAS.301..451G,Romero02,2010MNRAS.402..497G,2014Natur.515..376G} make possible a time-dependent analysis through the Fermi-LAT data even for those with $z>1$, while the available X-ray data suffers with gaps.


We study the gamma-ray activity of the flat spectrum radio quasars (FSRQs)~\citep{2011MNRAS.414.2674G} and BL Lacertae objects (BL Lacs)~\citep{1976ARA&A..14..173S} in the sample presenting more details for: OP 313, PKS 1454-354, GB6 J1040+0617 and TXS 0506+056, the ones within the selected blazars with higher statistics of the gamma-ray data available. With 9.5 years of Fermi-LAT data we obtain the gamma-ray light curves, duty cycles and luminosity during the major flares as well as the SED variation when the source moves from a quiescent state to a flaring state. Assuming the reported leptohadronic model~\citep{2015MNRAS.448.2412P}, we obtain the expected neutrino flux as a function of time (``neutrino lightcurve'') and compare it with the corresponding post-trial discovery potential flux of IceCube experiment~\citep{2018arXiv181107979I,2017ApJ...835..151A} for the different time bins considered. 
For the selected blazars we obtain the optimal flare duration to observe a neutrino event, similar to ones listed in the alert sample, in coincidence with an enhanced gamma-ray activity~\citep{2019MNRAS.489.4347O}.\\ On the other hand we show the importance of considering a duty cycle parameter whenever we link the possible neutrino emission with the electromagnetic spectrum for long-term observations~\citep{Tluczykont10,Patricelli14}. In this regard, we build a larger catalog of the most luminous blazars observed by Fermi-LAT and obtain the average duty cycle factor for this class of objects~\citep{2020arXiv_Sacahui} using it to correct the upper limits introduced recently~\citep{2018PhRvD..98f2003A,2019ApJ...871...41P,2019ApJ...880..103G} for the ratio between gamma-ray and VHE neutrino fluxes.
We assume these upper limits between the two fluxes to be constant with time, with the obvious consequence of being below the IceCube sensitivity limits during the quiescent states for the blazars studied.
We use a three-dimensional (3D) parameter space of gamma-ray luminosity/Duty Cycle/flare duration to compare the activity features. 
This analysis highlights the importance of increasing the size of neutrino telescopes and having a global neutrino network (GNN) to perform a multi-messenger observation with a significance greater than $3\sigma$ when observing a variable source with a neutrino fluence like the ones assumed here.
\section{Neutrino observations}
This analysis is performed after the VHE upward-going muon IC-170922A, reported by IceCube through a Gamma-ray Coordinates Network Circular on MJD 58018
(September 22, 2017;~\citep{2017GCN.21916....1K}) originating from a neutrino with energy 290 TeV, which has a high probably of having an astrophysical origin. Its best reconstructed position is right ascension (RA) 77:43+0:950:65 and declination (Dec) +5:72+0:500:30 (deg, J2000, 90$\%$ containment region: IceCube 
Collaboration 2018a). The other astrophysical track-like neutrino events considered in this work take into account the list of the 23 AMON alerts, the list of 
astrophysical muonic neutrinos reconstructed from the northern hemisphere~\citep{2016ApJ...833....3A} and the track-like events present in the last High Energy Starting Event (HESE) catalog~\citep{2017arXiv171001191I}. For each of these neutrino events we search for known blazar objects with a spatial correlation as described in section in the next section. 

\section{Source selection from the Fermi-LAT catalogs}
In this work we study the Fermi-LAT blazars that are spatially connected with IceCube track-like events. In particular, EHE and HESE events, from the Astrophysical Multimessenger Observatory Network (AMON) alert program~\citep{2013APh....45...56S,2017APh....92...30A} and muonic neutrino events above 200 TeV from~\citep{2016ApJ...833....3A} and~\citep{2017arXiv171001191I} (with $50\%$ containment error $\leq 1.5^{\circ}$) are considered. Additional events from~\citep{2017arXiv171001191I} satisfying the selection criteria are also included. We select blazars from the third Fermi Hard LAT (3FHL)~\citep{2017ApJS..232...18A} and the third Fermi Gamma-ray LAT 3FGL~\citep{Acero15} catalogs that are strictly spatially connected with these neutrino events. In particular we select only the objects with $|FermiLAT_{cen}-IceCube_{cen}|\leq1.3^{\circ}$ considering respectively the Fermi-LAT and the IceCube measured centroids. This condition leads to the position of reported blazars falling within $\sim5$ sq. deg. around the reconstructed astrophysical neutrino events.
The list of blazars that satisfy this condition are reported in Table~\ref{tab:sourcelist} with the corresponding distance, when known, and source types. 
For each of them we obtain the maximal gamma-ray luminosity reached during these 9.5 years considering different time binning and search for possible gamma-ray activity temporally correlated to an EHE neutrino event. We report in greater detail the cases of TXS 0506+056, OP 313, GB6 J1040+0617, PKS 1454-354, the ones within the sample which present fewer upper limits in the Fermi-LAT data, making possible a better study of the source variability. 
Although we use the spatial selection criteria for setting the list of sources and only afterward look for possible coincident gamma-ray flares, we also check the probability of a random spatial correlation between the known blazars (3FHL + 3FGL) and the neutrino events considered. Taking into account the total number of discrete BL Lacs and FSRQs present in the catalogues, the probability of a random spatial coincidence of one of these sources with the neutrino 50\% containment error of $1.5^{\circ}$ corresponds to a probability p=0.2. This number highlights the significance of the selected spatial trigger. Other spatial studies have also been reported in the literature with a weak correlation between the observed neutrinos and the gamma-ray activity of the sources observed by Fermi-LAT~\cite{Franckowiak:2020qrq,Giommi:2020hbx}.

\begin{table*}
\begin{center}
\caption{\label{source-list}Sample of blazars in spatial coincidence with selected IceCube with HESE and EHE $\nu_\mu$ events. With $z$ we report the redshift of the blazar.\\
1. \citep{2009yCat.2294....0A} 2. \citep{2012ApJS..203...21A} 3. \citep{2017Ap&SS.362..228P} 4. \citep{2015Ap&SS.357..141M} 5. \citep{2002A&A...386...97J} 6. \citep{2016A&A...590A..40B} 7. \citep{1989ApJS...69....1H} 8. \citep{Paiano18}} 
\label{tab:sourcelist}
\begin{tabular}{ l | l | l | l | l | l }
\hline\hline
S.no.  & Source Name & RA (deg.) & Dec. (deg.) & Source Class & \textit{z} \\ [0.5ex]
 \hline
1 & OP 313 & 197.649 & 32.351 & fsrq & $0.998 ^1$  \\
2 & SDSS J085410.16+275421.7 & 133.532 & 27.8826 & bll & $0.494 ^2$ \\
3 & 1RXS J064933.8-313914  & 102.386 & -31.6491 & bll & $\geq 0.563  ^3$ \\
4 & GB6 J1040+0617 & 160.147 & 6.302 & bll & $0.7351 ^4$ \\
5 & GB6 J1231+1421 & 187.866 & 14.368 & bll & $0.256 ^2$ \\
6 & PKS 1454-354 & 224.382 & -35.648 & fsrq & $1.424 ^5$ \\
7 & PMN J1505-3432 & 226.250 & -34.547 & bll & $1.554 ^6$ \\
8 & PMN J2227+0037 & 336.972 & 0.610 & bll & - \\
9 & PKS 2021-330 & 306.108 & -32.905 & fsrq & $1.47 ^7 $\\
10 & TXS 0506+056 & 77.3636 & 5.707 & bll & $0.336 ^8$ \\ [1ex]
\hline 
\end{tabular}
\end{center}
\end{table*}

\begin{table*}
\begin{center}
\caption{IceCube neutrino ($\nu_\mu$) events in spatial coincidence with the selected sources. The event types include: above 200 TeV event sample from the Northern Hemisphere ('NH-200T')~\citep{2016ApJ...833....3A, 2017arXiv171001191I}, track-like events from the HESE event selection~\citep{2017arXiv171001191I} and HESE and EHE alerts from AMON realtime alert system~\citep{2017APh....92...30A}. Event energies are cited without errors, and the deposited energy is listed as a lower limit to the neutrino energy where the best reconstructed energy of the neutrino is not known. Errors represent the 90\% C.L. in degrees for the case of NH-200T and the AMON alert events, while the median angular error is reported for the HESE sample. $D_{centroid}$ represents the distance between the best fit position of the neutrino event and the source position in the Fermi catalogs~\citep{2017ApJS..232...18A,Acero15}. 'Sglns' or Signalness, is the probability of the events to be of astrophysical origin, as calculated by IceCube.}
\label{tab:nueventlist}
\begin{tabular}{l|l|cc|cc|cc}
\hline\hline
Source                   & Event Type           & MJD                           & Energy (TeV)  & Sglns   & RA     & Dec   & $D_{centroid}$ \\ [0.5ex]
\hline
OP 313                   & NH-200T sample (\#17) & 56062.96                      & \textgreater 200 & 0.45 & $198.74^{+1.44}_{-1.09}$ & $31.96^{+0.81}_{-0.85}$     & 1.16        \\ [1ex]
SDSS J085410.16+275421.7 & NH-200T sample (\#32) & 57269.8                       & \textgreater 220 & 0.51 & $134^{+0.39}_{-0.58}$    &  $28^{+0.47}_{-0.47}$      & 0.48        \\   [1ex]
1RXS J064933.8-313914    & HESE sample (\#58)   & 56859.76                  & 52.6 & NA  & 102.1 $\pm 1.3$  & -32.4  $\pm 1.3$  & 0.80  \\ [0.5ex]
GB6 J1040+0617           & HESE sample (\#63)   & 57000.14                      & 97.4 & NA  & 160  $\pm 1.2$   & 6.5  $\pm 1.2$   & 0.25  \\ [0.5ex]
GB6 J1231+1421           & HESE sample (\#62)   & 56987.77                      & 75.8 & NA  & 187.9  $\pm 1.3$ & 13.3  $\pm 1.3$  & 1.07   \\ [0.5ex]
PKS 1454-354             & HESE alert           & 58405.49                     & 60 & 0.1  & 225.18  $\pm 1.23$ & -34.79  $\pm 1.23$  & 1.17 \\ [0.5ex]
PMN J1505-3432           & HESE alert           & 58405.49                      & 60 & 0.1 & 225.18  $\pm 1.23$ & -34.79  $\pm 1.23$  & 1.09  \\ [0.5ex]
PMN J2227+0037           & HESE sample (\#44)   & 56671.88                      & 84.6 & NA   & 336.71  $\pm 1.2$ & 0.04  $\pm 1.2$   & 0.63 \\ [0.5ex]
PKS 2021-330             & HESE alert           & 58507.15                      & NA  & 0.91 & 307.19  $\pm 1.23$ & -32.29  $\pm 1.23$  & 1.24 \\ [0.5ex]
TXS 0506+056             & EHE alert & 58018.87 & 290   & 0.56           & $77.43^{+0.95}_{-0.65}$  & $5.72^{+0.50}_{-0.30}$   & 0.09 \\ [0.5ex]
\hline
\end{tabular}
\end{center}
\end{table*}

\begin{figure}
\hspace*{-2.5cm}\includegraphics[scale=0.35]{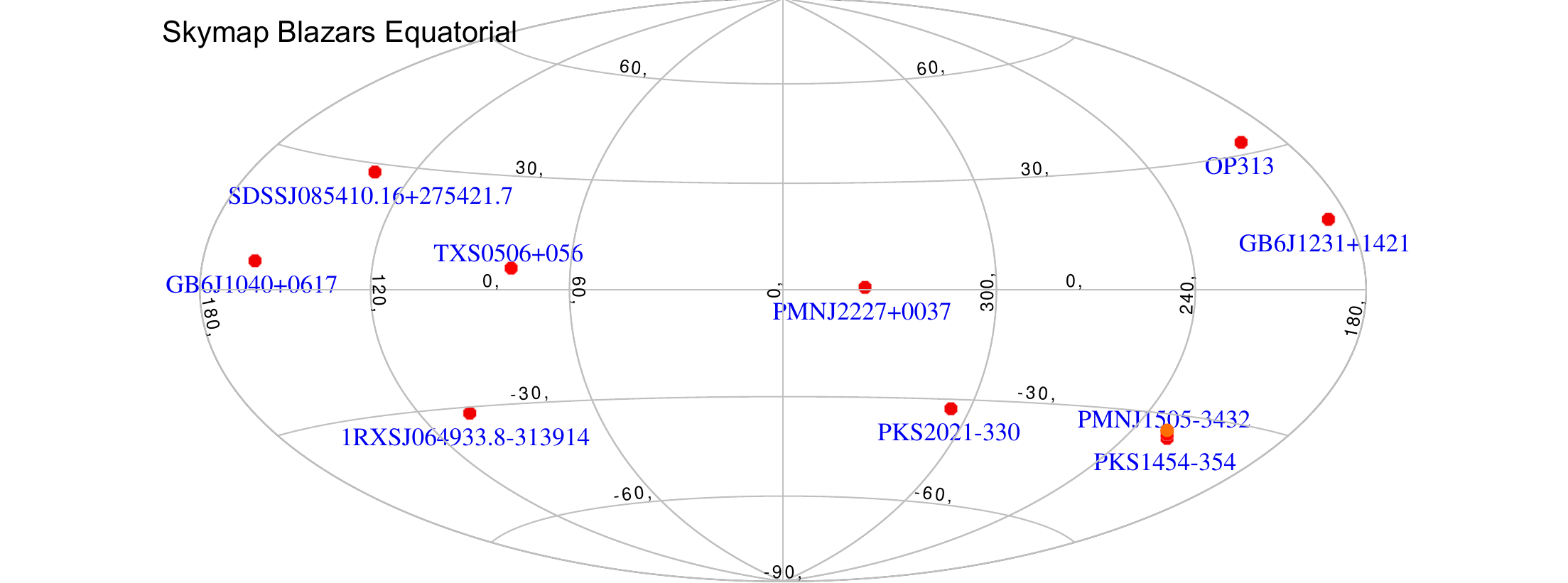}
\caption{\label{fig:Skympap-blazars-sample}Skymap in equatorial coordinates for the sample of Blazars reported in Table~\ref{tab:sourcelist}.}
\end{figure}

\subsection{TXS 0506+056}
TXS 0506+056 is a bright BL Lac type blazar, located at RA = $77.35^{\circ}$ and Dec. = $5.69^{\circ}$ (see Fig.~\ref{fig:Skympap-blazars-sample}). With a reported redshift of z=0.3365$\pm$ 0.0010~\citep{Paiano18}, it is the the most probable counterpart of the IceCube EHE neutrino event  IC-170922A~\citep{Kopper17}; spatially separated by $0.09^{\circ}$ from the reconstructed EHE event, within the $90\%$ containment region (event RA $77.43^{+0.95}_{-0.65}$ and Dec $5.72^{+0.50}_{-0.30}$). This blazar is found to have a synchrotron peak $(\nu^{S}_{peak})$ value below $10^{15}$ Hz (see Fig.~\ref{fig:nupeak-lum}) therefore classifiable between intermediate-frequency and high-frequency peaked blazar (IBL/HBL) as reported in~\citep{2018MNRAS.480..192P,2018ApJ-Murase}. The optical spectrum is typical of a Seyfert II galaxy~\citep{Paiano18} and supermassive black hole (SMBH) at the center with mass $M_{BH}\approx 3\times 10^{8}M_{\odot}$~\citep{Paiano18}. Recent studies have hinted that this blazar could be an FSRQ, masquerading as a BL Lac~\citep{10.1093/mnrasl/slz011}.



\subsection{OP 313}
OP313 is an FSRQ type blazar, located at RA = $197.619^{\circ}$ and  Dec. = $32.345^{\circ}$ (see Fig.~\ref{fig:Skympap-blazars-sample}), with a reported redshift of z=0.9980$\pm$ 0.0005~\citep{Hewett10}. As can be seen from Fig.~\ref{fig:nupeak-lum}, the $\nu^{S}_{peak}\sim 10^{13}$ Hz, and no estimation of the central SMBH mass is available. This FSRQ is spatially separated by $1.16^{\circ}$ from the reconstructed EHE event reported on May the $5^{th}$, 2012 by IceCube (event RA $198.74^{+1.44}_{-1.09}$ and Dec $31.96^{+0.81}_{-0.85}$), with a deposited energy of 200 TeV~\citep{2016ApJ...833....3A}. Exploring the gamma-ray emission, there is no temporal coincidence of the neutrino event with an increase of flux, but Fermi-LAT reported an increase in gamma-ray flux that began in April 2014, lasting for more than 3 months, and was up to 13 times its average flux~\citep{Sara14}.


\subsection{GB6 J1040+0617}

\begin{figure}
\centering
\hspace*{-0.5cm}\includegraphics[scale=0.25]{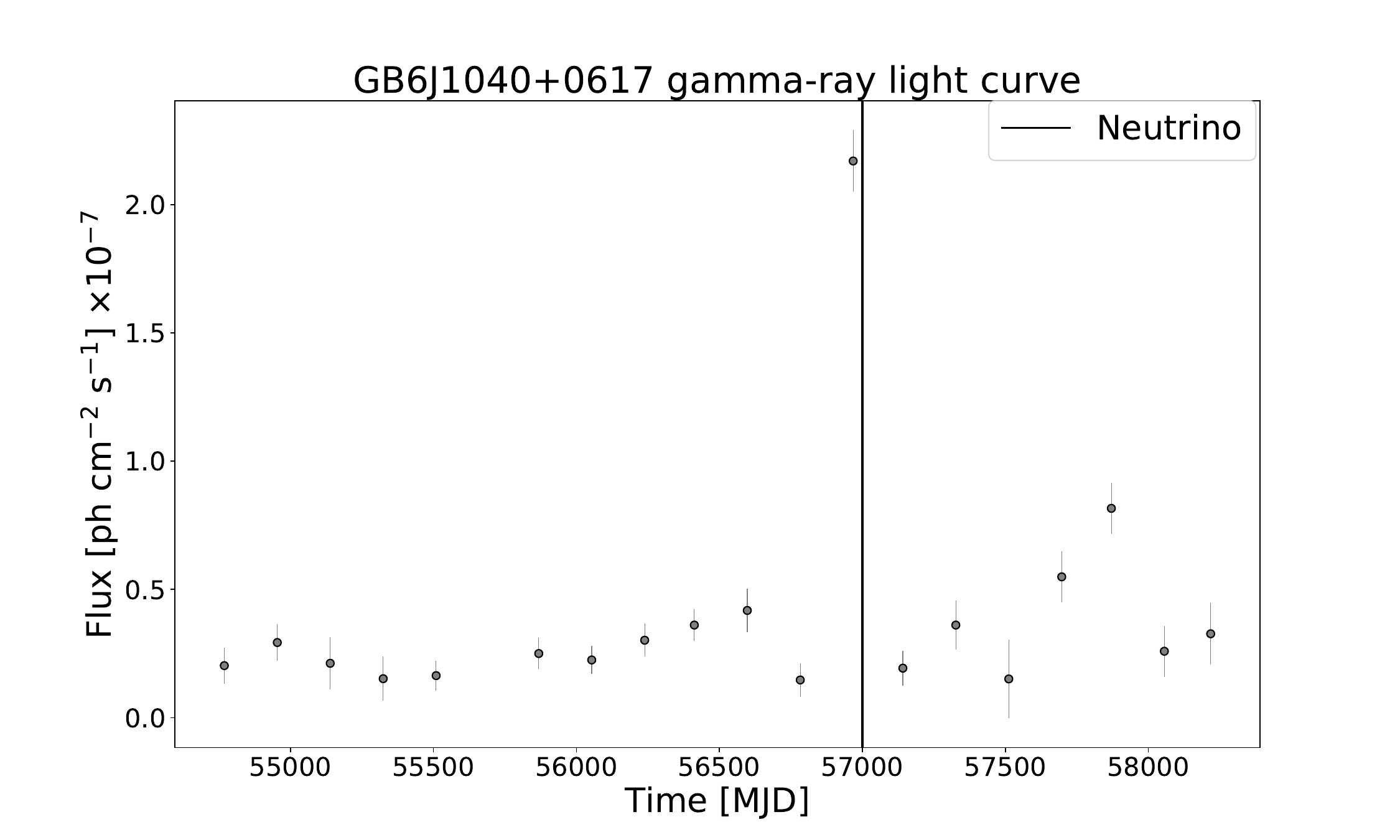}
\caption{Light curve (LC) of GB6 J1040+0617 obtained with Enrico software package~\citep{2013Enrico-soft_Sanchez} for 6 month time bins. 9.5 years of Fermi-LAT data between 0.1 - 300 GeV has been used. A power law spectrum is assumed and the EBL absorption is defined using~\citep{Franceschini08} model. The solid black line represents the time of the IceCube neutrino event in spatial coincidence with the source.}
\label{fig:LC_GB6}
\end{figure}

GB6 J1040+0617 is a BL Lac type blazar, located at RA = $160.147^{\circ}$, Dec. = $6.3023^{\circ}$, with a measured redshift of $z=0.7351$~\citep{gb6_redshift}. An EHE neutrino event (E $97.4^{+9.6}_{-9.6}$ TeV) detected in IceCube on 9th December 2014 at 03:26:04.000 UTC (MJD 57000.14310185; Fermi-MET 439788367), was reconstructed to be coincident with the direction of this blazar with a spatial separation of $0.25^{\circ}$ with 90\% C.L. (event RA $160$, Dec $6.5$, angular error $< 1.2^{\circ}$). As reported in~\citep{2019ApJ...880..103G}, the source shows increased gamma-ray activity for a period of $\sim 90$ days around the time of the neutrino detection. Another intensive period of activity is observed from the source for a much longer duration ($> 1$ year) starting MET 332035202, also accompanied by a hardening of the spectral index. The second smaller peak to the right of the neutrino event in the light curve (Fig.~\ref{fig:LC_GB6}) can be attributed to the emission from another source in the ROI, FSRQ 4C+06.41. This can be filtered by cutting on the lower energy range where uncertainties can be larger and plotting light curves for $E > 1$ GeV. GB6 J1040+0617 is a low synchrotron peaked blazar (LSP) with $\nu^{S}_{peak}\sim 6 \times 10^{13}$ Hz (Fig.~\ref{fig:nupeak-lum}). 


\subsection{PKS 1454-354}

\begin{figure}
\centering
\hspace*{-0.5cm}\includegraphics[scale=0.25]{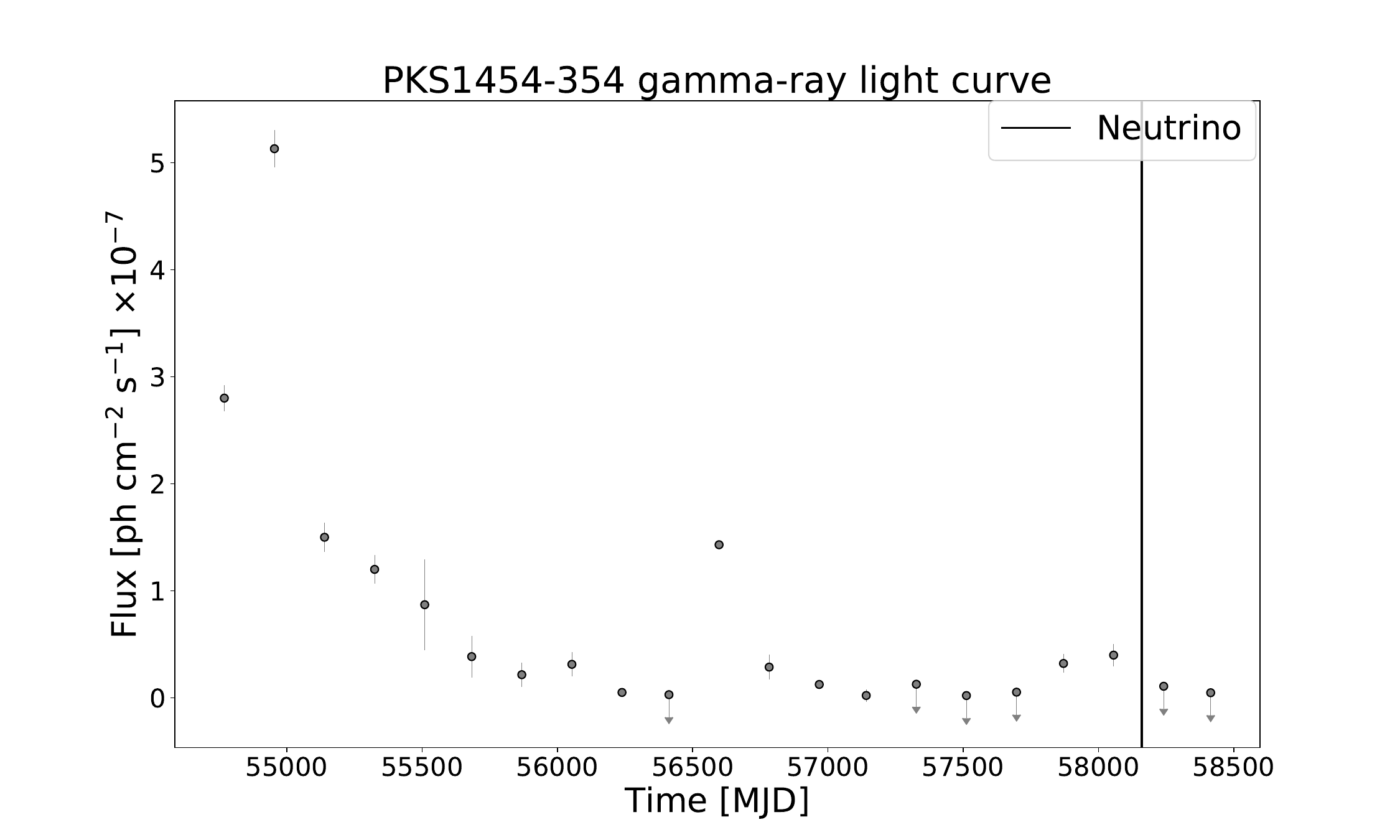}
\caption{Light curve (LC) of PKS 1454-354 obtained with Enrico software package~\citep{2013Enrico-soft_Sanchez} for 6 month time bins. 10+ years of Fermi-LAT data between 0.1 - 300 GeV has been used. A power law spectrum is assumed and the EBL absorption is defined using~\citep{Franceschini08} model. The solid black line represents the time of the IceCube neutrino event in spatial coincidence with the source.}
\label{fig:LC_PKS}
\end{figure}

PKS 1454-354 is an FSRQ type blazar. located at the coordinates RA = $224.382^{\circ}$, Dec. = $-35.6478^{\circ}$, in the Southern hemisphere. It has a measured redshift of $z=1.424$~\citep{jackson02}. It has been shown by~\citep{petrov05} to have a weak one-sided jet. The Fermi collaboration reported an increased $\gamma$-ray activity from the source in September of 2008~\citep{pks_fermi_paper}. The light curve of PKS 1454-354 (Fig.~\ref{fig:LC_PKS}) shows flaring activity from the source for an extended period of $\sim 1.5$ years, with a peak flux of $\sim 5.1 \times 10^{-7}$ $ph$ $cm^{-2}$ $s^{-1}$, higher than the average 10 year flux by a factor of almost 17. IceCube observed a HESE neutrino event (AMON HESE alert with coordinates RA = $225.184$, Dec. = $-34.792$) from the direction of this blazar on 14th October 2018 (IC181014A) with a reconstructed best-fit energy of $\sim 60$ TeV and a signalness (the likelihood of being an astrophysical event) of 0.38. It is an LSP type blazar ($\nu^{S}_{peak}\sim 1.12 \times 10^{13}$ Hz) as seen from Fig.~\ref{fig:nupeak-lum}, with a high $\gamma$-ray luminosity ($L_{\gamma} \simeq 3.8 \times 10^{47}$) in the Fermi energy range of 0.1-300 GeV. Another blazar from our sample, PMN J1505-3432, also lies in the ROI of this source.   




\section{Analysis of Fermi-LAT gamma-ray data}
\begin{figure}
\centering
\hspace*{-0.5cm}\includegraphics[width=0.52\textwidth]{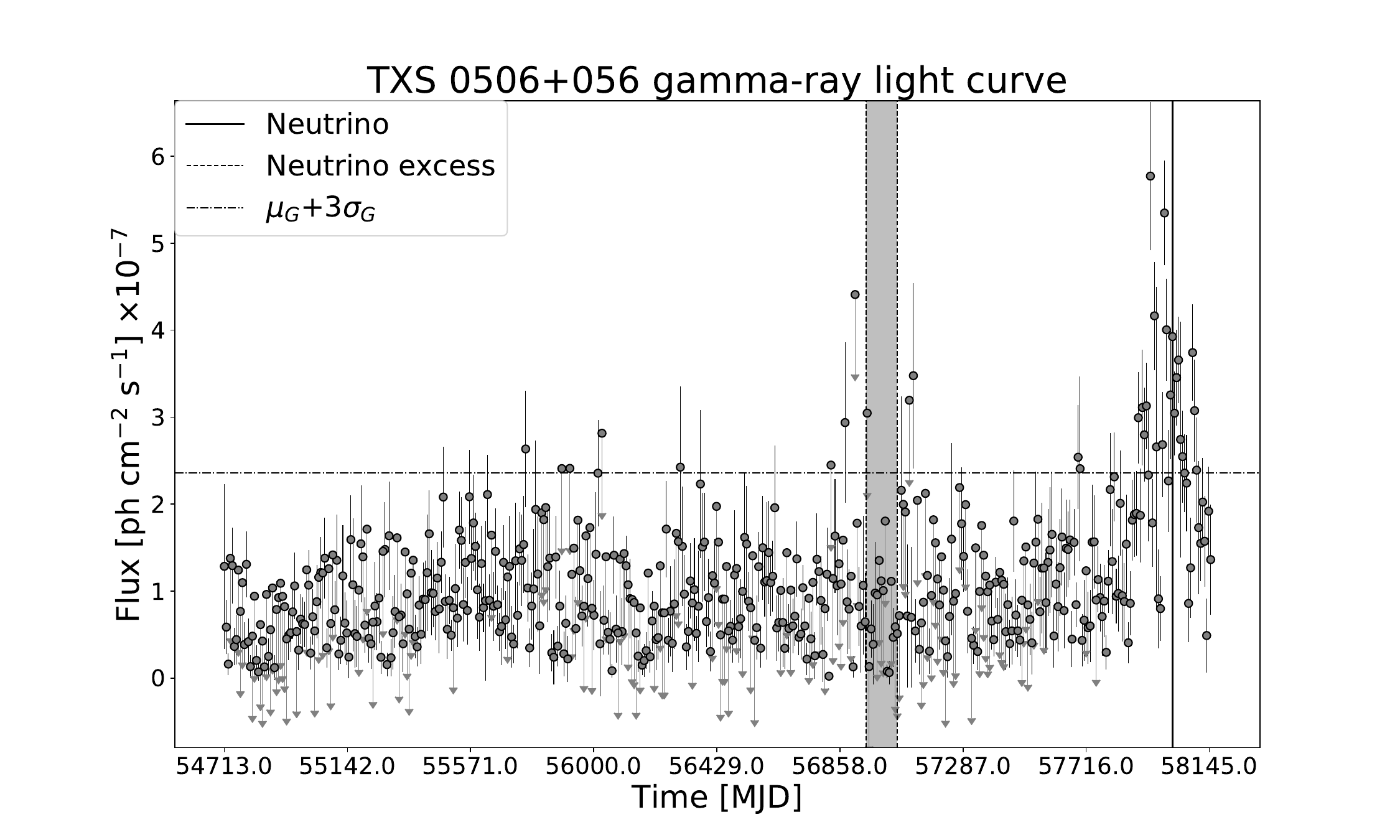}
\caption{\label{fig:TXS0506}Light Curve (LC) of TXS 0506+506 obtained with Enrico software package~\citep{2013Enrico-soft_Sanchez}. We assume a power-law behaviour for the spectral energy distribution (SED) of this source following the EBL model of~\citep{Franceschini08}. We report a weekly binned LC with the black solid vertical line indicating the time of the IC-170922A event and the shadowed region indicating the period of the 2014/2015 neutrino flare. The horizontal dashed-dotted line represents the $3\sigma$ deviation from the average flux.}
\end{figure}

\begin{figure}
\centering
\hspace*{-0.5cm}\includegraphics[width=0.52\textwidth]{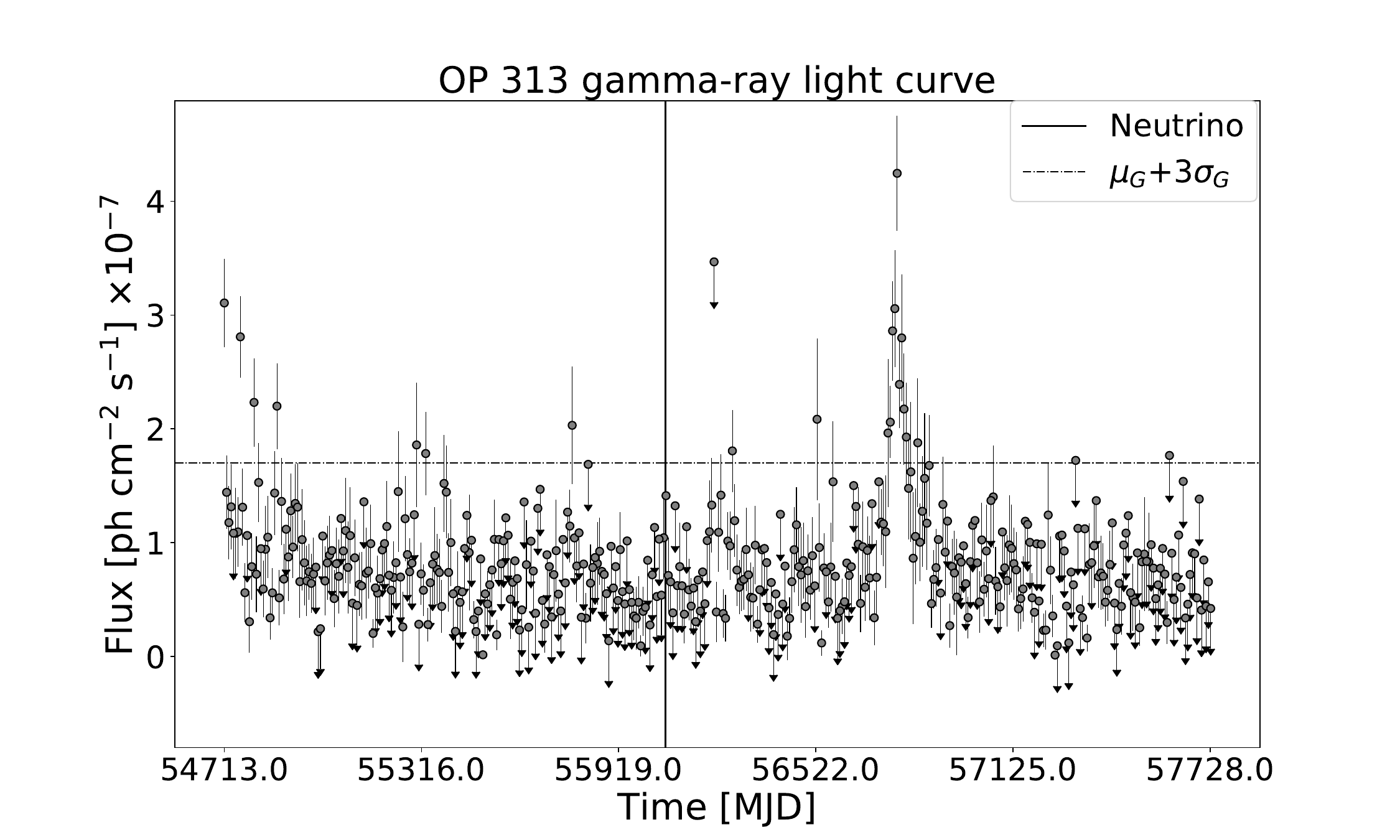}
\caption{\label{fig:OP313}Light curve (LC) of OP 313 obtained with Enrico software package~\citep{2013Enrico-soft_Sanchez}. Also for this source we describe the SED with a log parabola function and we follow the EBL model described by~\citep{Franceschini08}. We select a weekly binned flux with the black solid vertical line indicating the time of the observed EHE neutrino.}
\end{figure}

 We extracted data in the energy range 0.1 - 300 GeV within a $10^{\circ}$ Region of Interest (ROI) around each source. Data reduction was done using Enrico, a community-developed Python package to simplify Fermi-LAT analysis~\citep{Sanchez13}, using the LAT analysis software ScienceToolsv10r0p5 \footnote{http://fermi.gsfc.nasa.gov/ssc/data/analysis/software} and including all known gamma-ray sources reported in the third Fermi-LAT catalog~\citep{Acero15}, as well as the isotropic and Galactic diffuse emission components (iso\_P8R2\_SOURCE\_v6\_v06.txt and gll\_iem\_v06.fits\footnote{http://fermi.gsfc.nasa.gov/ssc/data/access/lat/BackgroundModels.html}), falling within the ROI.
We obtain the gamma-ray light curves with one week bins for the sources TXS 0506+056 and OP 313 from August 5, 2008 (MJD 54683) to February 5, 2018 (MJD 58154), see Figs.~\ref{fig:TXS0506},~\ref{fig:OP313}. While for the blazars GB6 J1040+0617 and PKS 1454-354, we use six month bin data between August 5, 2008 (MJD 54683) to June 5, 2018 (MJD 58275) and January 15, 2019 (MJD 58499) respectively. For the construction of the light curves, we adopt the spectra reported in the third Fermi-LAT catalog \citep{Acero15}. For TXS 0506+056, GB6 J1040+0617 and PKS 1454-354 we assume a power law spectrum while for OP 313 we assume a log parabola spectral shape.  We consider EBL absorption using~\citep{Franceschini08} as a reference model.
For each light curve we obtain upper limits when the test statistic was below 25 and with confidence level of $95\%$. With these light curves we calculate the corresponding source duty cycle, obtaining the GeV part of the SED and calculating the luminosity during the major observed flares (flares that are $1\sigma$ above the mean value). In the following subsections we treat all these aspects of the analysis in detail.

\subsection{Gamma-ray duty cycle} \label{sec:dc_sec}
\begin{figure}
\centering
\includegraphics[width=0.5\textwidth]{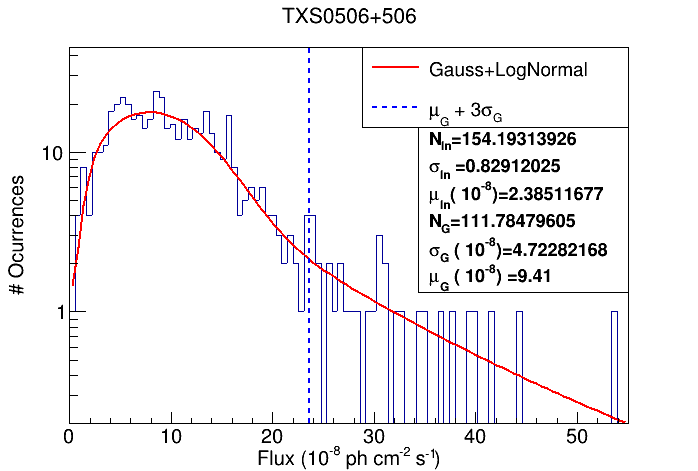}
\caption{\label{fig:fluxdist_TXS0506} In this plot we show the distribution of gamma-ray fluxes measured with Fermi-LAT telescope in 9.5 years in the energy range 0.1-300 GeV. Every observation corresponds to one week of data and the reported flux values on the Y-axis correspond to the average flux during the selected week. Overall the entire distribution is well described by a Gaussian+LogNormal distribution as done in~\citep{Tluczykont10}.}
\end{figure}

The duty cycle (DC) of a blazar can be defined as the fraction of time when the source is in a high flux state with respect to the total time of observation. The definition of a high flux state phase varies for different authors~\citep{Vercellone04,Krawczynski04}, and the DC can be expressed as,

\begin{equation}
DC=\frac{T_{fl}}{T_{fl}+T_{quies}}
\label{eq:DC}
\end{equation}
where $T_{fl}$ is the total time spent in a flaring or high flux phase and $T_{quies}$ is the total time spent in the quiescent or baseline flux state. Following~\citep{Abdo14,Patricelli14} the $DC$ can also be expressed as,

\begin{equation}
DC=\frac{\overline{F}-F_{bl}}{\langle F_{flare} \rangle + F_{bl}}
\label{eq:DCequ}
\end{equation}
where $\overline{F}$ is the average flux in the entire observation period and in the energy range used for the selection of the data, $F_{bl}$ is the baseline flux and $\langle F_{flare} \rangle$ is the average flux of flaring states. To infer the baseline flux we obtained the distribution of flux states for the $\sim9.5$ year Fermi-LAT observation period, considering 1 week bins, and as~\citep{Tluczykont10} reported for Mrk 421, it was best fit by a function consisting of a sum of a Gaussian ($f_{G}$) and a log-normal function ($f_{ln}$) with a likelihood integral fit. The flux distribution of TXS 0506+056 is reported in Fig.~\ref{fig:fluxdist_TXS0506} with the Gaussian + LogNormal fit. The mean of the Gaussian function represents the upper limit for the $F_{bl}$ and the log-normal function is associated to the flaring states~\citep{Tluczykont10}, hence the average flare flux can be expressed by:

\begin{equation}
\langle F_{flare} \rangle = \frac{\int_{F_{th}}^{F_{max}}xf_{ln}(x)dx}{\int_{F_{th}}^{F_{max}}f_{ln}(x)dx}
\label{eq:avg_flare}
\end{equation}

Since the Log-Normal function describes the flaring states, we choose as threshold flux $F_{th}$ = $F{bl}$ + $3\sigma_{G}$, and as maximum flux $F_{max}$ the highest flux observed in the light curves. 
The use of only one experiment to obtain the gamma-ray DC of the selected blazars minimizes the systematic errors for the obtained values. 
Here we report the DCs obtained for TXS 0506+056 and OP 313 using the weekly binning already shown in the light curve plots~\ref{fig:TXS0506},~\ref{fig:OP313}. 

Since we assume a leptohadronic emission~\citep{2015MNRAS.448.2412P} from these sources with the same emitting region for gamma-rays and neutrinos, we use the gamma-ray DC calculated trough Fermi-LAT data to also set the DC of neutrino emission.

It should be noted that the Gaussian + LogNormal fit of~\citep{Tluczykont10}, method I hereinafter, is only meaningful for sources that have discreet and well-defined flare(s)/active states and only few upper limits in the weekly or bi-weekly binned light curve obtained, which is not the case for a majority of sources in our sample. For this reason we report this method only for two sources: TXS 0506+506 and OP 313. To obtain a DC for our entire sample of blazars, we introduce an alternative method using a slightly modified approach from the one in~\citep{Vercellone04},
hereafter called method II. To estimate the DC, equation~\ref{eq:DC} is used. We computed the duration of the source flaring states which is defined as the number of consecutive time bins where the flux above a given threshold.
For this work, the mean flux is calculated including the upper limits, and the threshold for a flaring state is set to mean + $1\sigma$ ($\sigma$ = standard deviation). Only the bins whose error bars lie entirely above the threshold are considered for the calculation of active/flaring states. With this approach we obtain the DC for a sub-sample of 6 sources. Because of their low flux for these objects we obtained a two month bin LC for the 9.5y data and since all the time bins have the same length, the DC value was obtained by dividing the number of time bins in flaring state by the total number of bins (see Fig.~\ref{fig:lum-DC-time}). In Table~\ref{tab:lum_table} we report the average DC values of TXS0506+056 and OP313, the only two sources of the sample for which was possible to use method I with weekly time bins, as well as the duration of the major flares: the ones who presented the highest number of consecutive temporal bins with highest fluence.  


%
%

\subsection{Gamma-ray luminosity} \label{sec:lum_flare}

\begin{figure*}
\centering
\includegraphics[width=0.7\textwidth]{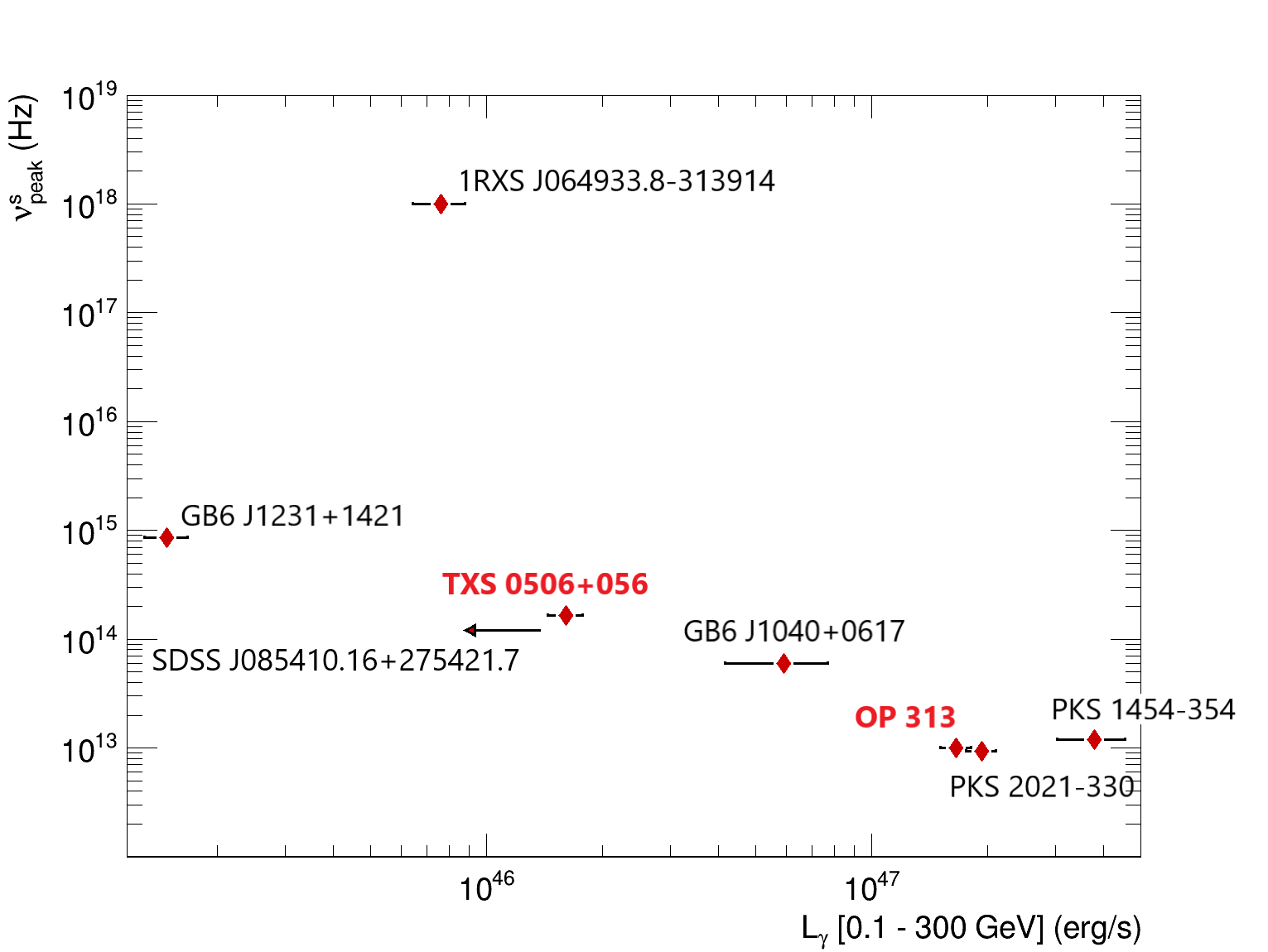}
\caption{\label{fig:nupeak-lum} Average gamma-ray luminosity (0.1 - 300 GeV) considering 9.5 years of Fermi-LAT data versus the synchrotron peak value from the Fermi-LAT 3FHL catalog for the sample of selected blazars. The extreme blazar 1RXS J064933.8-313914 is the only outlier in the anti-correlation trend followed by rest of the sources in the sample.}
\end{figure*}

We calculated isotropic gamma-ray luminosities for the sources (where redshift is available) in our sample. The calculation is based on the standard redshift-luminosity distance relation in a $\Lambda$CDM Universe. The cosmological parameters: $H_0$ = $67.8$, $\Omega_m$ = $0.308$, $\Omega_\lambda$ = $0.692$ are assumed for calculation, and redshifts of the sources are taken from Table~\ref{tab:sourcelist}. Wherever reported, the luminosities have been integrated in the energy range 0.1 - 300 GeV. 
Fig.~\ref{fig:nupeak-lum} reports the luminosities calculated using fluxes obtained in 9.5 - 10 years of Fermi-LAT data. The 3 FSRQs within the sample have higher luminosities than all BL Lacs, while 1RXS J064933.8-313914, the only extreme blazar in the sample, is the second least luminous. Luminosity during the most luminous flaring period of each source is reported in Table~\ref{tab:lum_table} and Fig.~\ref{fig:lum-DC-time}. The flaring periods for calculations of Table~\ref{tab:lum_table} are obtained using method I, while for Fig.~\ref{fig:lum-DC-time}, the flaring periods are obtained using method II. The luminosity of TXS 0506+056, accounting for the different periods of integration, is compatible with the one reported by~\citep{eaat1378}.\\
It is worth pointing out that the luminosities obtained in the above sections are strongly dependent on the redshift values of the sources through their luminosity distance.
For our selected sample, we also investigate possible common behaviour considering 9.5 years of Fermi-LAT data. When plotting the average gamma-ray luminosity vs. the synchrotron peak value, the combined sample of BL Lacs+FSRQs appears to follow an anti-correlation trend, proposed in the last decades for the integrated bolometric luminosity~\citep{2012MNRAS.420.2899G, 1999MNRAS.310..465G, Perlman:2000vm, Padovani:2003nw, 2005MNRAS.361..469M, Anton:2004iz, 2005A&A...434..385G, 2006A&A...445..441N, Raiteri:2015bvi, Padovani:2012hj, 2008MNRAS.387.1669G, Padovani07_review} as well as for the gamma-ray luminosity~\citep{fossati98, 2LAC_bolo_lum, blz_seq}, with the exception of the extreme blazar 1RXS J064933.8-313914 which appears as an outlier in Fig.~\ref{fig:nupeak-lum}. We analyze this trend to verify if our sample is dominated by high synchrotron peaked (HSP) blazars as seen in past studies correlating neutrinos and gamma rays~\citep{Padovani:2016wwn}. Apart from the already cited BL Lac 1RXS J064933.8-313914, this seems not to be the case, as indicated by comparing the trend of Fig.~\ref{fig:nupeak-lum} with a similar blazar sequence study reported in~\citep{2020arXiv_Sacahui} using a much larger sample of sources.

\begin{table*}
\begin{center}
\caption{\label{tab:lum_table}Average photon fluxes, isotropic luminosities during their most luminous flares and their duration, calculated using method I. Average values of the duty cycles obtained, when possible, following~\citep{Tluczykont10} criterion.}
\begin{tabular}{ l | c | c | c | c | c | c}
\hline\hline
Source &  Av. flare flux & Luminosity & Lum. Error &  Flare duration &Avg. DC  \\
&  (ph cm$^{-2}$ s$^{-1}$)& (erg s$^{-1}$) & erg s$^{-1}$ &(weeks) &  \\[0.5ex]
\hline 
TXS 0506+056 & $3.38 \times 10^{-7}$  & $5.33 \times 10^{46}$ & $6.12 \times 10^{45}$  & 20 &  $\sim 23\%$\\
OP 313  & $2.54 \times 10^{-7}$ & $6.06 \times 10^{47}$ & $1.79 \times 10^{47}$ & 8  & $\sim 21\%$\\ 
GB6 J1040+0617 & $7.48 \times 10^{-7}$ & $1.03 \times 10^{48}$ & $4.58 \times 10^{46}$ & 4  & -\\
GB6 J1231+1421 & $4.60 \times 10^{-7}$& $5.24 \times 10^{45}$ & $1.94 \times 10^{45}$  & 4 & -\\
PKS 1454-354 & $3.45 \times 10^{-7}$ & $2.43 \times 10^{48}$ & $3.90 \times 10^{47}$  & 24 & - \\
PKS 2021-330 & $6.32 \times 10^{-8}$ & $4.82 \times 10^{47}$ & $8.08 \times 10^{46}$ & 8 & - \\

\hline 
\end{tabular}
\end{center}
\end{table*}

\section{Spectral Energy Distributions}
\begin{figure}
\centering
\hspace*{-0.5cm}
\includegraphics[width=0.55\textwidth]{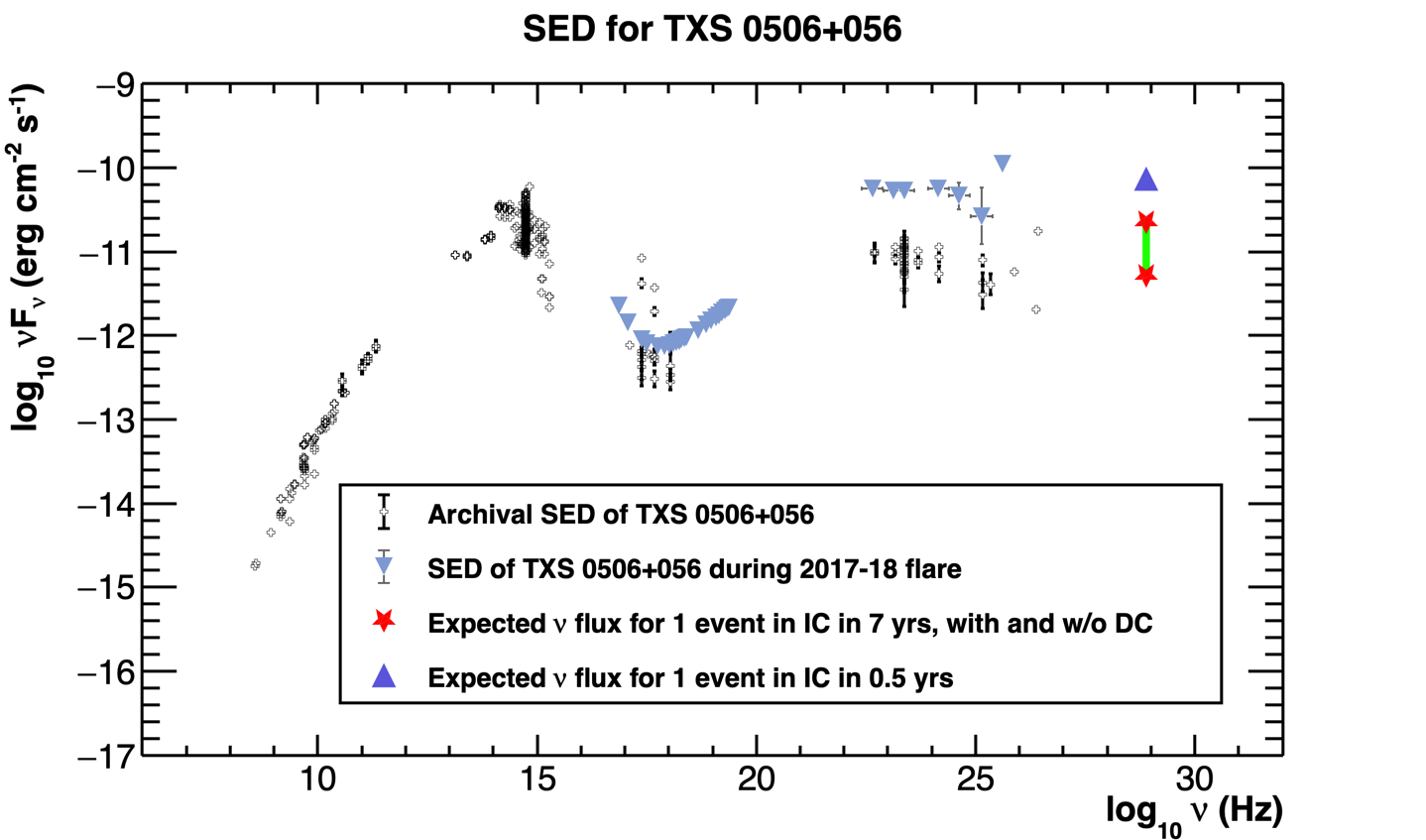}
\caption{\label{fig:TXS0506sed} Multi-messenger SED of TXS 0506+056. With the black points we report the electromagnetic SED obtained using the open access multi-wavelength archival data from the ASDC SED Builder Tool of Italian Space Agency (ASI)~\citep{stratta2011asdc}. With the carolina triangle we report the Fermi-LAT gamma-ray data and the X-ray data from SWIFT-XRT, NuSTAR during the flaring period. We use the violet triangle to indicate the expected neutrino flux to obtain a astrophysical neutrino event considering 6 months of Icecube data taking, while the the two red stars, separated by the green region, indicate the neutrino flux variation when considering 7 years of IceCube data taking and a source duty cycle varying from $0\%$ to the value of $23\%$ reported in Table~\ref{tab:lum_table}.}
\end{figure}

\begin{figure}
\centering
\hspace*{-0.5cm}
\includegraphics[width=0.55\textwidth]{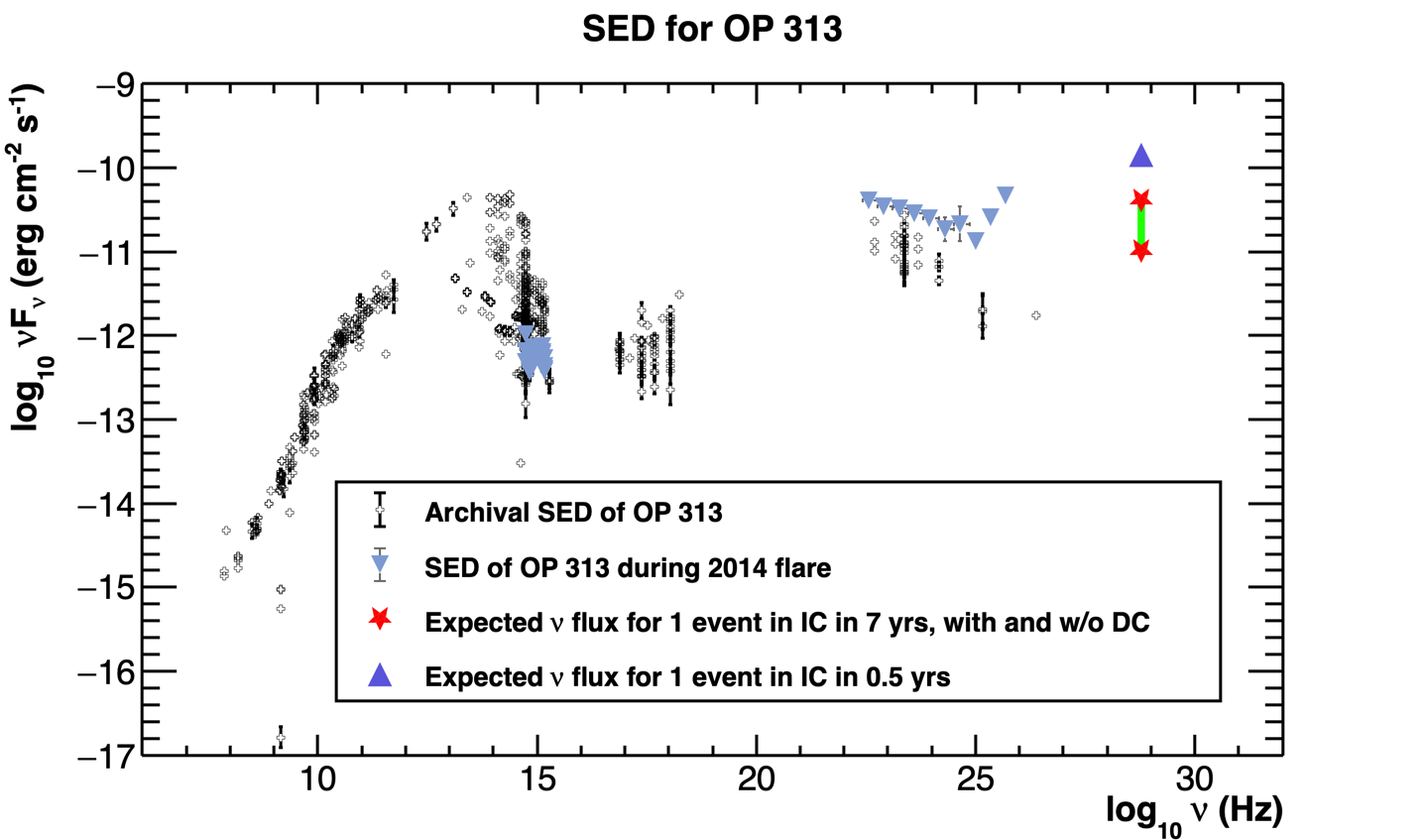}
\caption{\label{fig:OP313sed} Multi-messenger SED of OP 313. With the black points we report the electromagnetic SED obtained using the open access multi-wavelength archival data from the ASDC SED Builder Tool of Italian Space Agency (ASI)~\citep{stratta2011asdc}. With the carolina triangle we report the Fermi-LAT gamma-ray data and the UV to soft X-ray data during the flaring period. We use the violet triangle to indicate the expected neutrino flux to obtain an astrophysical neutrino event considering 6 months of Icecube data taking, while the the two red stars, separated by the green region, indicate the neutrino flux variation when considering 7 years of IceCube data taking and the a source duty cycle varying from $0\%$ to the value of $21\%$ reported in Table~\ref{tab:lum_table}.}
\end{figure}

\begin{figure}
\centering
\hspace*{-0.5cm}
\includegraphics[width=0.55\textwidth]{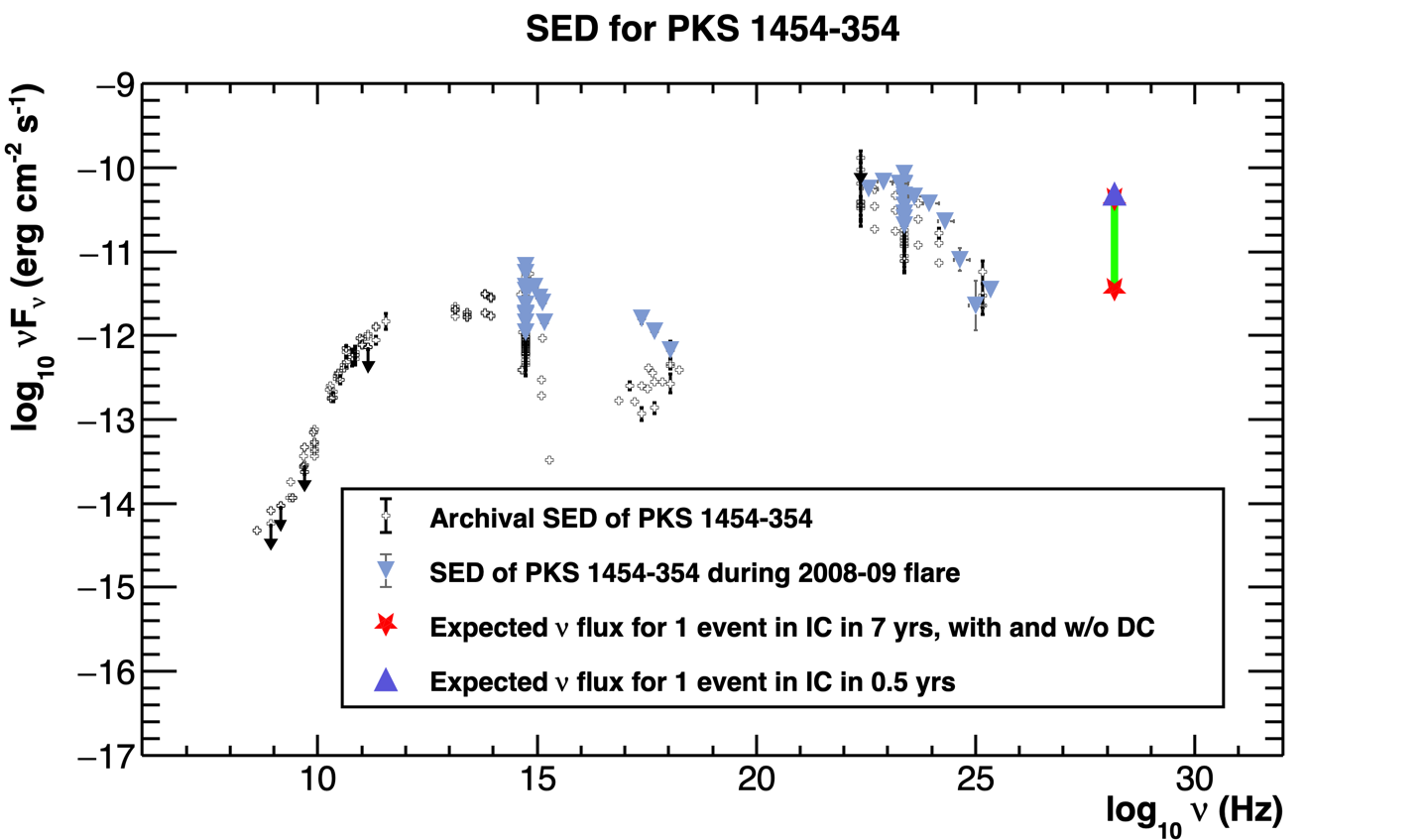}
\caption{\label{fig:PKS1454sed} Multi-messenger SED of PKS1454-354 . With the black points we report the electromagnetic SED obtained using the open access multi-wavelength archival data from the ASDC SED Builder Tool of Italian Space Agency (ASI)~\citep{stratta2011asdc}. The carolina triangles indicate the Fermi-LAT gamma-ray data, optical-UV and X-ray data during the flaring period. The violet triangle is used to indicate the expected neutrino flux to obtain a astrophysical neutrino event considering 6 months
of IceCube data taking, while the the two red stars, separated by the green region, indicate the neutrino flux variation considering 7 years of IceCube data taking and a source duty cycle variation between 0\% and the value of 8.33\% obtained with the method described in \citep{Vercellone04}.}
\end{figure}

\begin{figure}
\centering
\hspace*{-0.5cm}
\includegraphics[width=0.55\textwidth]{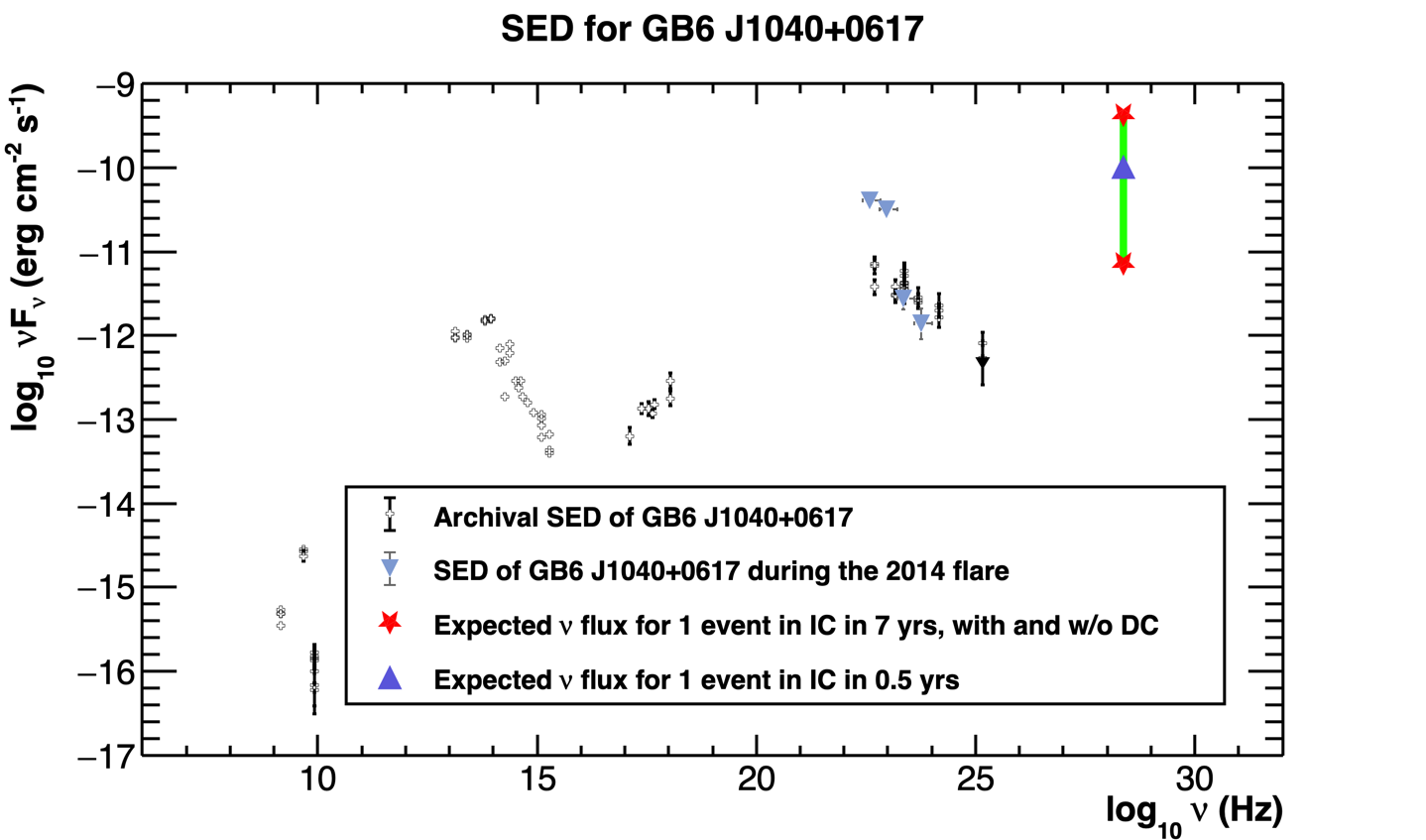}
\caption{\label{fig:GB6J1040sed} Multi-messenger SED of GB6J1040+0617. With the black points we report the electromagnetic SED obtained using the open access multi-wavelength archival data from the ASDC SED Builder Tool of Italian Space Agency (ASI)~\citep{stratta2011asdc}. The carolina triangles indicate the Fermi-LAT gamma-ray data during the flaring period. The violet triangle is used to indicate the expected neutrino flux to obtain a astrophysical neutrino event considering 6 months of IceCube data taking, while the the two red stars, separated by the green region, indicate the neutrino flux variation considering 7 years of IceCube data taking and a source duty cycle variation between 0\% and the value of 1.72\% obtained considering the metod of \citep{Vercellone04}.}
\end{figure}

In this work we also consider the broadband spectral energy distribution (SED) of the selected blazars to link the electromagnetic spectrum with the neutrino observations. Here we report the SED of the blazars TXS 0506+056 (Fig.~\ref{fig:TXS0506sed}), OP 313 (Fig.~\ref{fig:OP313sed}), GB6 J1040+0617 (Fig.~\ref{fig:GB6J1040sed}), PKS 1454-354 (Fig.~\ref{fig:PKS1454sed}).
For the entire SED range we use the open-access multi-wavelength archival data from the ASDC SED Builder Tool of Italian Space Agency (ASI)~\citep{stratta2011asdc} during the 9.5 years of Fermi-LAT data taking, considering the following additional instruments: KVA~\citep{Lindfors2016}, UVOT and XRT onboard the Neil Gehrels Swift observatory~\citep{Roming:2005hv}, NuSTAR~\citep{Harrison:2013md}.
Data obtained at various wavelengths during the most significant gamma-ray flare for each of the above sources is highlighted over the archival data. For the gamma-ray energy range in particular, the flaring state data points are obtained from our analysis of Fermi-LAT data. We compare the gamma-ray spectral features for these four cases to the corresponding neutrino flux needed to obtain an observable EHE muonic neutrino event in IceCube from the position of the blazar considered. The differential neutrino fluxes in the SED are obtained for a single event in IceCube from the source using the IceCube effective area for tracks at the declination of the respective sources and the observational time of 6 months and 7.5 years. Fluxes are evaluated at the energy of the spatially correlated neutrino events, with a spectral assumption of $E^{-2}$. The source duty cycle is also factored in while calculating the neutrino flux expectation over the 7.5 years of observation. We use the values of duty cycle reported in Table~\ref{tab:lum_table} for TXS 0506+056 and OP 313 (obtained with method I) and the ones visible in Fig~\ref{fig:lum-DC-time} for GB6 J1040+0617 and PKS 1454-354 (obtained with method II) to set the range of possible neutrino fluxes when looking at such a long observational time. Since we calculate this flux value just considering the observed neutrino event to be spatially correlated, no $K_{\nu\gamma}$ factors are considered.



\section{Expected neutrino emission} \label{sec:model}
Both BL Lacs and FSRQs can be considered VHE neutrino emitter candidates since their jets can accelerate protons up to ultra high energies (UHE)~\citep{2015APh....70...54F}. When the density of the synchrotron emission from the electrons in the jet exceeds the density of gas, the accelerating protons can make the following three interactions: \textit{i}) $p\gamma\rightarrow p{\gamma}^{'}$, inverse Compton scattering; \textit{ii}) $p+\gamma \rightarrow p e^{+} e^{-}$, electron positron pair production; and \textit{iii)} $p\gamma \rightarrow N + k\pi$, photomeson production. When the energy of target photons in the rest frame of protons exceeds $m_{\pi}c^{2}\times$(1 + $m_{\pi}$/$2m_{p}$)$\simeq 145$ MeV the photomeson production starts and neutrino production starts to be effective through the decay of $\pi^{+}$ and $\pi^{-}$.\\
In this work we follow the~\citep{2015MNRAS.448.2412P} approach to estimate the expected neutrino flux for the sources in our sample. In particular we consider the gamma-ray data recorded by Fermi-LAT observatory between 1 GeV and 300 GeV during the flaring period, to be mainly produced by the synchrotron emission of pion cascade products inside the jets of the blazars considered. In this context the target for the photo-hadronic interactions of ultra-relativistic protons can be composed by environmental blazar photons like: X-ray synchrotron photons emitted by electrons, broad line region photons or photons from the jet sheath. This leptohadronic approach from~\citep{2015MNRAS.448.2412P} used to evaluate the expected neutrino flux is not testable in the X-ray energy range during the flaring time for most of the reported sources due to a lack of data. However, for the case of TXS 0506+056, there are simultaneous X-ray and gamma-ray observations~\citep{eaat1378, 2018ApJ_MAGIC} available around the time of arrival of IC-170922A and a modeling of the broadband SED of the source in a time window around the gamma-ray flare~\citep{2019NatAs...3...88G,2018ApJ-Murase}. Constraints imposed by the X-ray data strongly suggest that a single-zone emission model, where the dominant gamma-ray component in the high energy hump of the SED is related to hadronic processes, cannot reconcile the X-ray observations and the expected neutrino flux~\citep{2018ApJ-Murase, 2019MNRAS_Zech, 2019NatAs...3...88G}. Two-zone emission models, however, like the spine-sheath model by~\citep{2005A&A...432..401G, Tavecchio_2014}, where a slow moving sheath surrounds the fast-moving ultra-relativistic central spine in a structured jet, have been successfully employed to explain the multi-messenger observations of TXS 0506+056 at various epochs~\citep{2019ApJ...874L..29R, 2018ApJ_MAGIC, 2019NatAs...3...88G}.\\ To relieve the tension between our single-zone model of choice ~\citep{2015MNRAS.448.2412P} and the X-ray observations for the case of TXS 0506+056, we adopt an approach favoring a major leptonic component for the observed gamma-rays. In particular, the LMBB2b scenario described in~\citep{2018ApJ-Murase}, which considers the Inverse Compton upscattering by synchrotron electrons as the dominant component in the high energy hump of the EM SED, is implemented to compute the neutrino SED from TXS 0506+056 during its 2017-18 gamma-ray flare. \\
Different proportionality constants $K_{\nu\gamma}$ were considered to link the emission measured between 1 GeV and 300 GeV ($L_{\gamma}$) and the corresponding neutrino expectations between 100 TeV to 1 PeV ($L_{\nu}$):

\begin{equation}
K_{\nu\gamma} = \frac{L_{\nu}}{L_{\gamma}}
\label{eq:frac_nu_gamma}
\end{equation}

The calculated neutrino fluxes $F_{\nu_{\mu} + \overline{\nu}_{\mu}}$ reported in Fig.~\ref{fig:OP-TXS-nuLC} are obtained considering two different values for $K_{\nu\gamma}$, 1 and 0.5.
$K_{\nu\gamma}=1$ is reported as a ``baseline'' case, with a $1:1$ ratio between the flux of photons from secondary particle emission and the neutrinos from charged pion decays, while $K_{\nu\gamma}=0.5$ is introduced as a  ``benchmark'' case where the DC of a large sample Blazars~\citep{2020arXiv_Sacahui} is taken into account as a weighting parameter and applied to the limit of $K_{\nu\gamma}=0.1$ obtained recently \citep{2016PhRvL.117x1101A,2017ApJ...835...45A,2018PhRvD..98f2003A,2019ApJ...871...41P,2019ApJ...880..103G}. In fact, the value of $K_{\nu\gamma}=0.5$ is comes from the ratio $0.1/0.2$, and convolutes the limit on neutrino emission from blazars, obtained through the 3FHL catalog, with the average DC obtained considering a large blazar sample. For the case of TXS 0506+056, the external Compton model adopted from~\citep{2018ApJ-Murase} predicts a difference of an order of magnitude between the neutrino differential flux and the gamma-ray differential flux in the Fermi-LAT range, equivalent to $K_{\nu\gamma}=0.1$. Thus, the two K-values considered for TXS 0506+056 are $K_{\nu\gamma}=0.1, 0.5$, taking into account the DC as well.
In the following part we will refer to $K_{\nu\gamma}$ using simply a $K$. 
In this context we assume low values of opacity ($\tau_{\gamma\gamma}$) caused by the interaction of hundred GeV photons with the Broad Line Region (BLR) photons, similar to what is reported in~\citep{PhysRevLett.116.071101,10.1093/mnras/227.2.403}. 

\subsection{Expected neutrino flares observability}
As explained in the previous section, we obtain the expected neutrino flux from the Fermi-LAT data considering different proportionality constants $K$; while for TXS 0506+056 we follow the model of~\citep{2018ApJ-Murase}, for the other blazars we consider the gamma-rays to be emitted by the products of pion decays~\citep{2015MNRAS.448.2412P}. Although originally applied to BL Lacs, this treatment can also be valid for FSRQs with a high proton injection luminosity ($L_p$), considering that external photon fields can also contribute to the neutrino production~\citep{2014PhRvD..90b3007M}. We apply these scenarios for different time bins, following the EBL parametrization of~\citep{Franceschini08} and assuming an SED described by a power law with exponential cutoff for the Fermi-LAT data. 
However, for the corresponding neutrino fluxes we consider a power law SED with $\alpha = 2$ without discriminating between the models assumed for the different sources or between the possible different spectral behaviours between the flaring states and the quiescent states. Differentiating the blazar description with time would prevent the construction of a ``neutrino lightcurve'' over such a long observational period, even though a state by state spectral index can be considered in a future work.\\
The time-binned neutrino fluxes obtained for the entire period of Fermi-LAT data taking are compared to the discovery potential of IceCube telescope for the same time period. The position of the source as well as the energy peak of expected neutrinos are also considered while obtaining the discovery potential flux.The expected neutrino SED is estimated on a source by source basis and the discovery potential is evaluated at this energy. Specifically, for TXS 0506+056, the reported discovery potentials correspond to the minimum (160 TeV) and maximum (120 PeV) of the neutrino SED in the LMBB2b model of~\cite{2018ApJ-Murase}, which has been applied to TXS 0506+056 in our work. For OP 313 and PKS 1454-354, both FSRQs, we obtain the expected neutrino energy peak following Eqn. 24 of~\citep{2014PhRvD..90b3007M}, with the Broad Line Region (BLR) emission - in the UV to soft X-ray band, acting as the target photons for neutrino production in the $\mathcal{O}(100)$ TeV - $\mathcal{O}(100)$ PeV range. In particular we use the data available for these sources in the above band during their gamma-ray flare (as highlighted in Figs.~\ref{fig:PKS1454sed} and~\ref{fig:OP313sed}) to calculate the neutrino SED peaks. For OP 313, the peak is obtained at E $ \sim160$ PeV using the UV data at ($\nu \sim 10^{15}$ Hz), while for PKS 1454-354, the corresponding peak at E $ \sim200$ TeV is obtained with the soft X-ray data at ($\nu \sim 10^{18}$ Hz).\\For GB6 J1040+0617 however, this procedure cannot be followed due to lack of any simultaneous flaring state data in the UV to X-ray range. Also, shifting the target photon energy towards the synchrotron peak results in a neutrino SED (obtained with the procedure outlined in~\cite{Padovani:2015mba}, since it is a BL Lac) lying outside the range of IceCube sensitivity. So the discovery potential for GB6 J1040+0617 is instead evaluated at the energy of the neutrino event associated with this source (E $ \sim97$ TeV). The differential discovery potentials for the Northern sky reported by IceCube in~\citep{disc_new} are used in the calculations, except for PKS 1454-354, for which the results from~\citep{2017ApJ...835..151A} for the Southern sky are used.





\begin{figure*}
\centering
\includegraphics[width=1.0\textwidth]{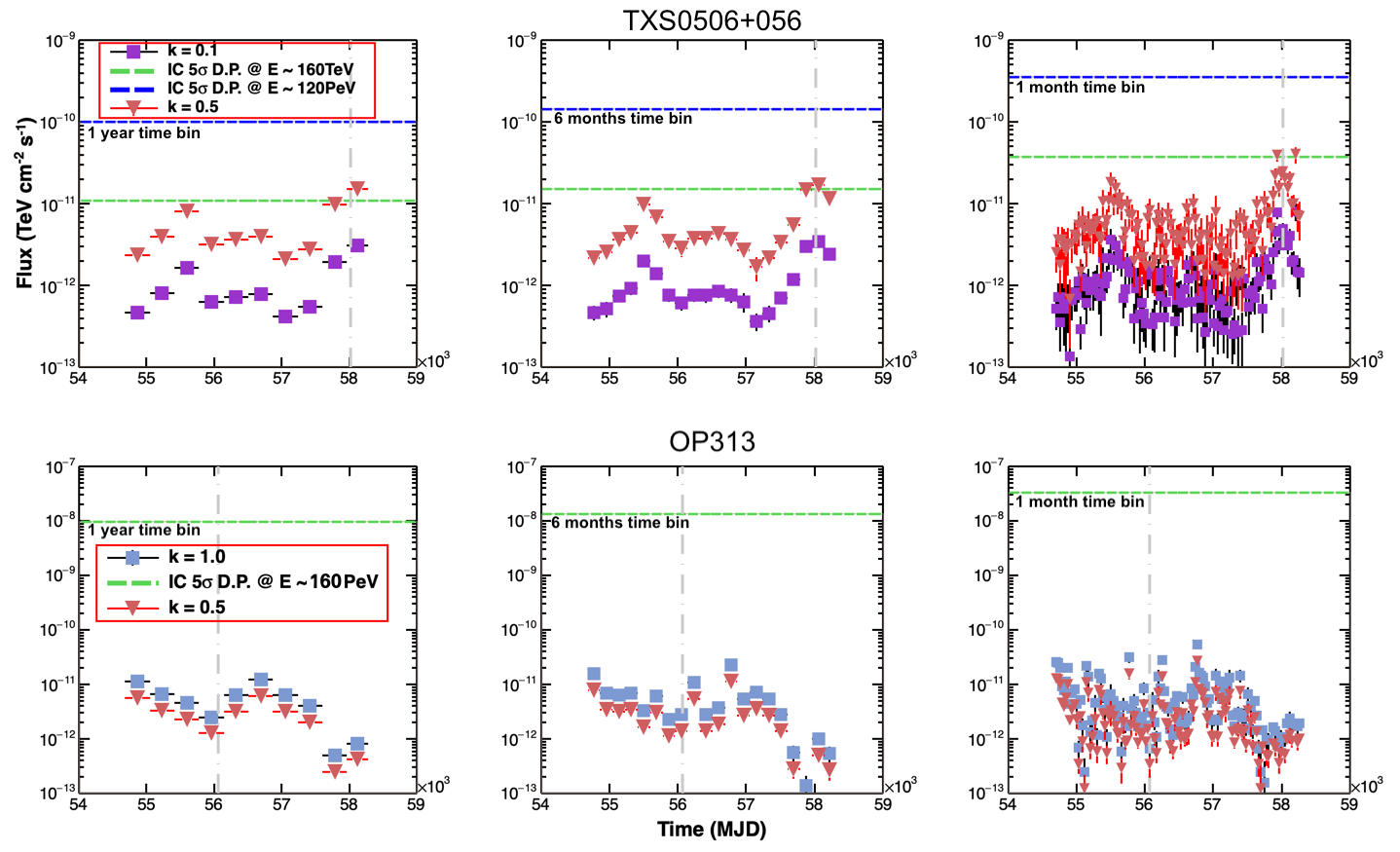}
\caption{\label{fig:OP-TXS-nuLC} Neutrino LCs of OP 313 and TXS 0506+056. On the top panel, the neutrino LCs expected from TXS 0506+056 considering 1 year, 6 months and 1 month time binning (left to right), while on the bottom panel, the same neutrino light curves for OP 313. The green and blue dashed lines correspond to the IceCube discovery potential with 50\% confidence level from~\citep{disc_new} for the neutrino energy peak expected following the criteria described in the text and scaled for the time bin considered. The corresponding neutrino energy is reported in the legend. For TXS 0506+056, the green and blue discovery potentials refer to the minimum and maximum neutrino energy expected following~\citep{2018ApJ-Murase}. For OP313 the green line corresponds to the discovery potential at the energy value obtained from~\citep{2014PhRvD..90b3007M}. The light curves of TXS 0506+056 are computed assuming $K_{\nu\gamma}=0.1, 0.5$ while for OP 313 $K_{\nu\gamma}=1, 0.5$ is assumed. The grey dashed-dotted line corresponds to the time of the neutrino event used for the spatial trigger of the blazar.}
\end{figure*}

\begin{figure*}
\centering
\includegraphics[width=1.0\textwidth]{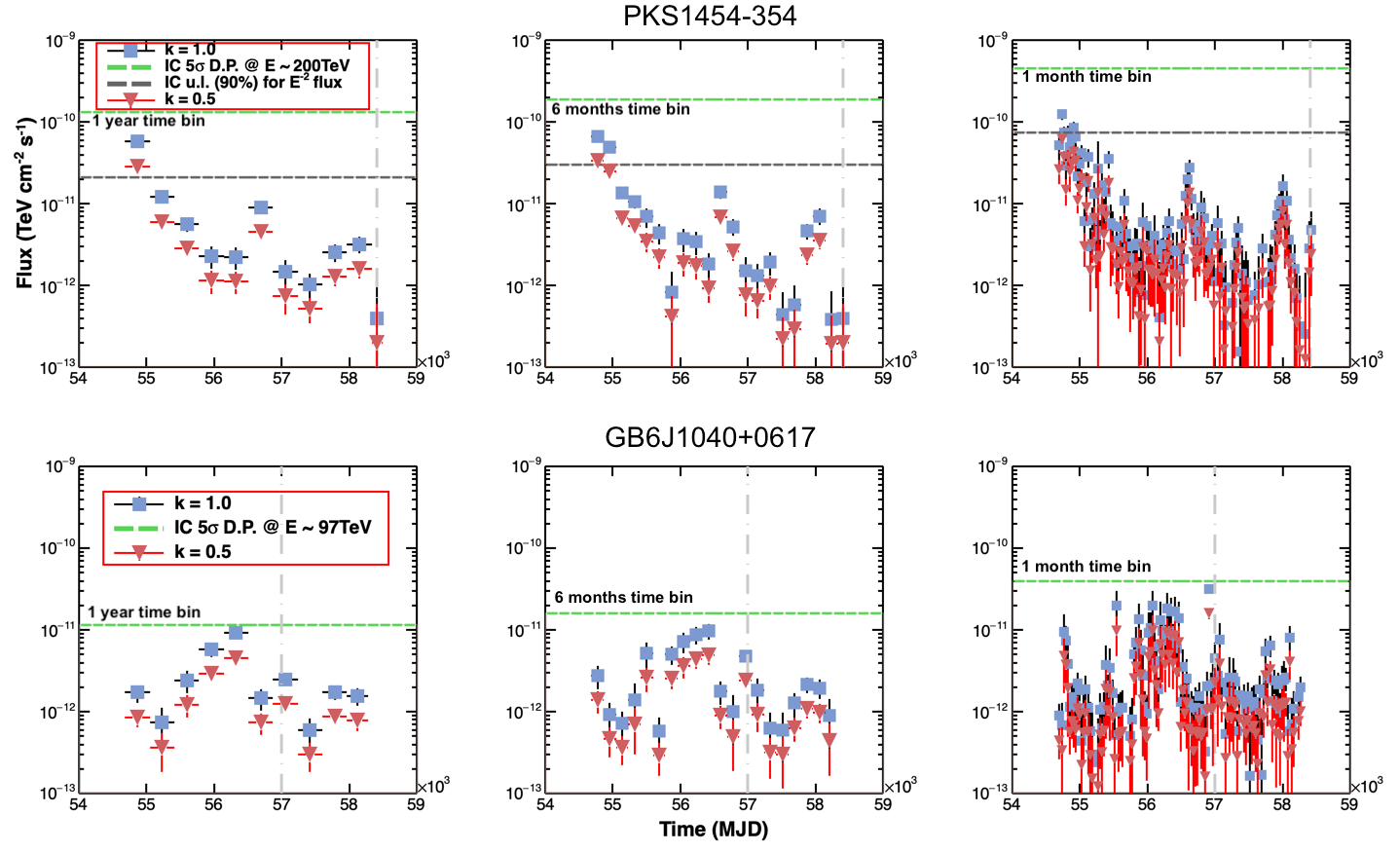}
\caption{\label{fig:PKS-GB6-nuLC} Neutrino LCs of the blazars PKS 1454-354 and GB6 J1040+0617. On the top panel, the neutrino LCs expected from PKS 1454-354 considering 1 year, 6 months and 1 month time binning (left to right), while on the bottom panel, the same neutrino light curves for GB6 J1040+0617. The green dashed lines correspond to the IceCube discovery potential with 50\% confidence level from~\citep{disc_new} for GB6 J1040+0617 and from~\citep{2017ApJ...835..151A} for PKS 1454-354, at the neutrino energy peak expected following the criteria described in the text and scaled with the time bin considered. The corresponding neutrino energy is reported in the legend. For PKS1454-354, the green line corresponds to the discovery potential at the energy value obtained from~\citep{2014PhRvD..90b3007M} and the grey dashed line corresponds to the flux upper limit imposed by IceCube for this source~\citep{Aartsen:2019fau}. For GB6 J1040+0617, the green line corresponds to the discovery potential calculated at the energy of the spatially correlated neutrino event. The light curves are computed for the two cases $K_{\nu\gamma}=1$ and $K_{\nu\gamma}=0.5$. The grey dashed-dotted line corresponds to the time of the neutrino event used for the spatial trigger of the blazar.}
\end{figure*}

This analysis is aimed at finding the optimal activity period needed to observe a possible EHE track-like event correlating the emission of photons of $\mathcal{O}(10^2)$ GeV with the neutrinos of $\mathcal{O}(10^2)$ TeV in the jet of the blazar. In Fig.~\ref{fig:OP-TXS-nuLC} and Fig.~\ref{fig:PKS-GB6-nuLC} we report the two cases of $K$ = 1.0 and $K$ = 0.5 for three of the blazars that we find in the sample: OP 313, GB6 J1040+0617 and PKS 1454-354, while for TXS 0506+056 we consider $K$ = 0.1 and $K$ = 0.5.


\section{Discussion}
\begin{figure*}
\centering
\includegraphics[width=1.0\textwidth]{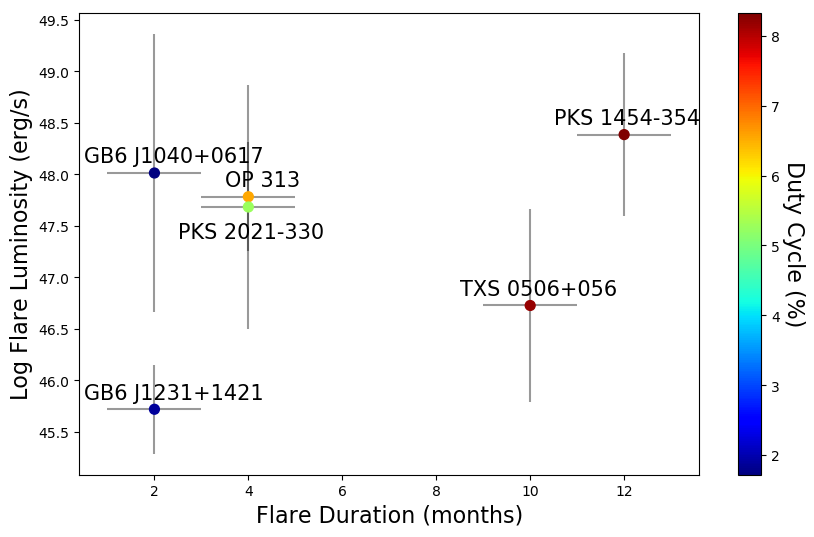}
\caption{\label{fig:lum-DC-time} In this plot, for each source we report the gamma-ray luminosity during the most luminous flare vs the average duration of the gamma-ray flare. The color scale indicates the minimum duty cycle obtained with method II as explained in Section~\ref{sec:dc_sec}. The values are obtained in the energy range 0.1-300 GeV using 9.5 years of Fermi-LAT data. Not all the blazars listed in Table~\ref{tab:sourcelist} are reported here due to the low statistics of photons at high energy and due to the difficulties in obtaining a lower limit for the duty cycle.}
\end{figure*}

For the studies presented in this work we select potential VHE emitters (being part of 3FHL or 3FGL catalogs) spatially located within 1.3$^{\circ}$ of the centroid of a HESE or EHE muonic neutrino event. A fraction of these neutrino events are included in the list of IceCube alerts, while others were reconstructed before the alert program. Some events that are not part of the alerts are also included. With this search we produce a sample of 10 sources reported in Table~\ref{tab:sourcelist}. 
Considering that the selected strict spatial correlation with the blazars reported in the 3FGL and 3FHL catalog can produce for each neutrino event a random coincidence with less than 0.2 sources, we look for additional gamma-ray and neutrino correlations verifying if the gamma-ray activity can explain the neutrino observations.
For each source we look first for possible temporal coincidence of gamma-ray flare (average + $3\sigma$) considering a weekly time binning during the first 9.5 years of Fermi-LAT observations, with the selected neutrino events. 
From Fig.~\ref{fig:lum-DC-time} we can see the level of gamma-ray activity for the blazar sample when we use a bi-monthly time scale.
The FSRQ PKS 1454-354 is seen to be the object with a higher level of gamma-ray activity and a longer flare duration, however only TXS 0506+056 is found to have a significant and long gamma-ray flare in coincidence with the reconstructed EHE neutrino event (of $22^{nd}$ September 2017).\\
The limit reported in literature for $K_{{\nu}{\gamma}}$, already discussed 
in Section \ref{sec:model}, makes the detection of neutrino events during the blazar quiescent period difficult and highlights the importance of obtaining a DC parameter for this class of sources whenever a a long term multi-messenger study, relating neutrinos and gamma-rays, is considered.
Taking into account the statistics of gamma-ray events above 0.1 GeV for TXS 0506+056 and OP313 we obtain the DC applying the approach described by~\citep{Tluczykont10} when using the weekly time bin. Instead, for the other blazars in our sample, with less statistics on collected gamma-rays, we use a slightly modified method compared to~\citep{Vercellone04} with monthly time bins. 
The DC of blazars in our sample resulted to be in accordance with the average DC obtained for a larger sample of the 40 most luminous blazars in the 3FHL catalog~\citep{2020arXiv_Sacahui}.\\ 
Here we obtain also a possible ``neutrino lightcurve'' derived from the gamma-ray lightcurve, following two interesting approaches: the simple application of the leptohadronic model explained in Section~\ref{sec:model}, the ``baseline'' case, and an optimized ``benchmark'' case. For the ``benchmark'' case we take into account the reported upper limit for the neutrino/gamma differential flux ratio of the known blazars and the average DC from a larger blazar sample.\\
We compute these ``neutrino lightcurves'' for different time bins and compare them with the flux needed to obtain a $5\sigma$ discovery in $50\%$ of IceCube equivalent pseudo-experiments, in the same amount of time. This analysis shows that the minimal flaring time needed to observe with IceCube telescope an integrated neutrino flux above 100 TeV, is of the order of a few months, for a very luminous BL-Lac. Looking at these lightcurves in detail, a one month period seems to be enough for TXS 0506+056 to cross the IceCube discovery potential threshold while 
none of the other blazars presents enough neutrino flux to be detected in the time bin of 1 and 6 months, and 1 year. This can suggest: a different emission process for the gamma-rays observed, a major gamma-ray absorption by the EBL due to the distance of the sources, or a casual spatial coincidence with the observed astrophysical neutrinos.\\ 
Additionally, we present the study of the spectral features considering the total time of Fermi-LAT data taking and the period of major gamma-ray flare for TXS 0506+056, OP 313, PKS 1454-354 and GB6 J1040-0617. In the Figs.~\ref{fig:TXS0506sed}, \ref{fig:OP313sed}, \ref{fig:PKS1454sed}, \ref{fig:GB6J1040sed} we report the level of expected differential neutrino flux to observe an EHE event and we compare it with the (0.1 - 300 GeV) differential gamma-ray flux for the total Fermi-LAT period and for the flaring times. The level of neutrino flux needed for the detection of 1 EHE event is seen to be higher than the gamma-ray differential flux during the flaring periods, however still compatible with the possibility to describe the photons of $\mathcal{O}(10^2)$ GeV emitted at least by OP 313, PKS 1454-354 and GB6 J1040-0617, through the model described in~\citep{2015MNRAS.448.2412P}. In Fig.~\ref{fig:lum-DC-time} we characterize the observed gamma-ray flares and highlight the peculiar activity of TXS 0506+056 and PKS 1454-354; within this phase space, a red point (high DC) lying in top right region of the figure, would represent a good candidate to be observed through high energy neutrinos.


\section{Results and Conclusions}
The first step of this work was a creation of a sample of blazars with enhanced gamma-ray emission spatially and temporally correlated with the astrophysical-like neutrino events reconstructed by IceCube telescope. 
With the built sample, the goal was to compare the time-dependent gamma-ray activity of the BL-Lac 
TXS 0506+056 with the other potential VHE neutrino sources. Two important aspects were 
examined, considering that these kind of high-energy astrophysical emitters spend a sizeable part of their life 
in an ``off'' state: the Duty Cycle and the minimum flaring period required to be observed through a 
kilometric neutrino telescope. Other 9 blazars, fulfilling the criteria of 
strict spatial correlation with an EHE neutrino event, although without temporal coincidence, were analyzed. 
Other sources, like OP 313, PKS 1454-354 and GB6 J1040-0617,  with a high average flux and a relatively high DC are good candidates to constrain the hadronic emission from the blazars. 
Here we show that the minimum hadronic flare duration needed to obtain a multi-messenger observation (with $> 100$ TeV neutrinos and $> 1$ GeV gamma-rays) when considering a very luminous BL Lac (the ones among the brightest 3\% of the 3FHL catalog), should be of the order of few months. We realize this computing the ``neutrino lightcurve'', modeling the electromagnetic SED through leptohadronic approaches. We also emphasize the importance of calculating the DC values for these variable and powerful objects.
The DC value at the level of $\sim 20\%$ or below for the blazars in this sample justify the fact that we don't have, at the moment, multiple neutrino events above 100 TeV from known blazars and that IC-170922A, in coincidence with the gamma-ray flare of TXS 0506+056, is the only multi-messenger observation. Our study did not attempt to obtain the contribution of the entire class of blazars to the diffuse neutrino flux measured by IceCube, however the introduction of a DC would change the upper limit of blazar contribution to the astrophysical IceCube flux. 
In the near future, an interesting similar study, connecting VHE neutrinos with X-ray emission, would be the computation of blazars' DC with the incoming  data of the new X-ray telescope eROSITA~\citep{2010HEAD...11.4804B}.
On the other hand, this work confirms the fact that on a short time interval a multi-messenger observation of blazars with $ > 3\sigma$ confidence level can be obtained only through bigger neutrino telescopes and possibly with a global neutrino network comprising several multi-$km^3$ sized detectors spread around the world. This will soon be possible with IceCube and the upcoming KM3NeT~\citep{2016JPhG...43h4001A} and Baikal-GVD~\citep{2014NIMPA.742...82A} telescopes. 
\section*{Acknowledgements}
The authors would also like to thank Barbara Patricelli, Kohta Murase and Foteini Oikonomou for their helpful suggestions and discussions regarding the hadronic activity estimation of blazars. The authors acknowledge the financial support of the funding agencies: {JRS} acknowledges DIGI-USAC for financial support by grant 4.8.63.4.44 and DGAPA-UNAM IG101320.; Istituto Nazionale di Fisica Nucleare (INFN), Ministero dell’Istruzione, dell’ Universit\`a e della Ricerca (MIUR), PRIN 2017 program (Grant NAT-NET 2017W4HA7S) Italy.

\section*{Data Availability}
The data and code essential for reproducibility of the results of this article are publicly available on the Zenodo platform through the DOI: \href{https://doi.org/10.5281/zenodo.4731071}{10.5281/zenodo.4731071}.

\bibliographystyle{mnras}
\bibliography{BibRod}

\end{document}